\DocumentMetadata{}
\documentclass[sigplan,nonacm]{acmart}

\usepackage[]{hyperref}
\usepackage{dashrule}
\usepackage{amsmath}
\usepackage{bm}
\usepackage{subfig}
\usepackage{enumitem}
\usepackage{xspace}
\usepackage{multirow}
\usepackage{listings}
\usepackage{xcolor}
\usepackage{soul}
\usepackage{fancyvrb}
\usepackage{amsfonts}
\usepackage{xcolor,colortbl}
\usepackage[breakable]{tcolorbox}
\usepackage[subtle]{savetrees}
\usepackage[export]{adjustbox}
\setlist[itemize]{leftmargin=*}
\usepackage{algorithm}
\usepackage{algpseudocode}
\usepackage{booktabs}
\usepackage{pifont}

\definecolor{bg}{rgb}{0.95,0.95,0.95}

\begin{document}
\newcommand{\todo}[1]{\textcolor{blue}{TODO: #1}}
\newcommand{\txs}[1]{{\small\texttt{#1}}}
\newcommand{\txsfig}[1]{{\footnotesize\texttt{#1}}}
\newcommand{\swatch}[1]{\textcolor{#1}{$\blacksquare$}}
\definecolor{deepred}{HTML}{EA2027}
\definecolor{deepblue}{HTML}{1B1464}
\definecolor{forestgreen}{HTML}{006266}
\definecolor{ultraviolet}{HTML}{341f97}
\definecolor{seafoam}{HTML}{1289A7}
\definecolor{radiantyellow}{HTML}{F79F1F}
\definecolor{burntorange}{HTML}{d35400}
\definecolor{darkgrey}{HTML}{273c75}
\definecolor{pastelblue}{HTML}{E3F4F4}
\definecolor{lightgrey}{HTML}{D3D3D3}
\definecolor{merchantmarineblue}{HTML}{0652DD}

\definecolor{keywordcolor}{HTML}{1289A7}
\definecolor{apicolor}{HTML}{341f97} 
\definecolor{functioncolor}{HTML}{1B1464}
\definecolor{variablecolor}{HTML}{0652DD}
\definecolor{argumentcolor}{HTML}{006266}
\definecolor{commentcolor}{HTML}{273c75}

\definecolor{plotred}{HTML}{d62728}
\definecolor{plotblue}{HTML}{1f77b4}
\definecolor{plotgreen}{HTML}{2ca02c}
\definecolor{plotorange}{HTML}{ff7f0e}

\newcommand{\tool}{\textsc{Shem}\xspace{}}
\newcommand{\specabbrev}{XXXSpec}
\newcommand{\uacs}{novel analog computing systems} 

\newcommand{\optalgo}{mapping optimization}
\newcommand{\ark}{Ark}
\newcommand{\normaldist}{\mathcal{N}}
\newcommand{\codein}[1]{{\texttt{\small #1}}}
\newcommand{\mathin}[1]{{\small\textbf{$#1$}}}
\newcommand{\rulein}[1]{{\small\textsc{#1}}}

\newcommand{\jax}{JAX}
\newcommand{\equinox}{Equinox}

\newcommand{\Rsq}[0]{R^2}
\newcommand{\ddt}[1]{\frac{d#1}{dt}}
\newcommand{\loss}{\mathcal{L}}
\newcommand{\vecstatevar}{\mathbf{x}}
\newcommand{\param}{\theta}
\newcommand{\vecparam}{\boldsymbol{\param}}
\newcommand{\vecadjoint}{\mathbf{a}}
\newcommand{\vecadjointgrad}{\mathbf{u}}

\newcommand{\vecmismatch}{\boldsymbol{\delta}}
\newcommand{\vectransient}{\boldsymbol{\xi}}

\newcommand{\cnn}{CNN}
\newcommand{\obc}{OBC}
\newcommand{\tln}{TLN}
\newcommand{\nn}{NNs}

\newcommand{\modelabbr}{AIS}
\newcommand{\noiseamp}{\delta}
\newcommand{\temporalweight}{c}
\newcommand{\weightbit}{q}

\newcommand{\tightbullet}{\noindent}
\newcommand{\proseheading}[1]{\tightbullet{\textbf{{#1}}}}
\newcommand{\prh}[1]{\noindent\textbf{#1}}
\newcommand{\pri}[1]{\noindent\textit{#1}}
\newcommand{\bulletkeyword}[1]{\textbf{#1}}
\newcommand{\sectionbullet}[2]{\textbf{#1[{\small\texttt{#2}}].}}
\newcommand{\mathbullet}[2]{\textbf{#1[{\small$#2$}].}}

\newcommand{\itoo}{\overline{I2O}}
\newcommand{\db}{\mathrm{dB}}

\newcommand{\allcplweight}{k}
\newcommand{\cplweight}{\allcplweight_{ij}}
\newcommand{\lockweight}{S_l}
\newcommand{\cplscale}{S_c}
\newcommand{\snp}{SnP}

\newcommand{\makeset}[1]{\mathbf{#1}}
\newcommand{\makefunc}[1]{\mathrm{#1}}
\newcommand{\bit}[1]{b_{#1}}
\newcommand{\chl}{c}
\newcommand{\centerchls}{\makeset{C}}

\newcommand{\PUFrawfn}{\makefunc{D}}
\newcommand{\PUFbinfn}{\makefunc{P}}
\newcommand{\bfneighbor}{\makefunc{bf}}
\newcommand{\sign}{\makefunc{is\_positive}}

\renewcommand{\algorithmicrequire}{\textbf{Input:}}
\renewcommand{\algorithmicensure}{\textbf{Output:}}
\newcommand{\algkw}[1]{\textbf{#1}\xspace{}}
\newcommand{\algfnprose}[1]{\texttt{#1}}
\newcommand{\algin}{\algkw{in}}
\newcommand{\algmatchprod}{\algkw{LookUpProdRule}}
\newcommand{\algapp}{\textit{append}}
\newcommand{\algnodes}{\algkw{NodesOf}}
\newcommand{\algedges}{\algkw{EdgesOf}}
\newcommand{\algnt}{\algkw{NodeType}}
\newcommand{\alget}{\algkw{EdgeType}}

\newcommand{\prodrule}{\phi}
\newcommand{\varmap}{\mathcal{M}_x}
\newcommand{\attrmap}{\mathcal{M}_\theta}

\definecolor{swanwhite}{HTML}{f7f1e3}
\definecolor{dirtybrown}{HTML}{aaa69d}
\tcbset{genericStyle/.style={
    colback=white,
    left=10pt,
    right=10pt,
    boxsep=0pt,
    breakable,
}}

\tcbset{research-question/.style={
    genericStyle,
    sharp corners,
    colframe=dirtybrown,
    boxrule=0.5pt,
    colback=swanwhite!75,
}}

\newtcolorbox[]{research-question}[1][] {research-question,#1}

\newcommand{\nc}{${\mathsf{NC}}$}
\newcommand{\bo}{${\mathsf{BO}}$}
\newcommand{\hebbian}{${\mathsf{HEBB}}$}
\newcommand{\baseline}{${\mathsf{BASE}}$}
\newcommand{\autoparallel}{${\mathsf{AP}}$}

\newcommand{\beforeedit}[1]{\textcolor{red}{\st{#1}}}
\newcommand{\afteredit}[1]{\textcolor{forestgreen}{#1}}

\title{Generative Models on  Analog Hardware with Dynamics}

\author{Yu-Neng Wang}
\email{wynwyn@stanford.edu}
\affiliation{%
  \institution{Stanford University}
  \country{USA}
}
\author{Sara Achour}
\email{sachour@stanford.edu}
\affiliation{%
  \institution{Stanford University}
  \country{USA}
} 

\begin{abstract}

Analog hardware platforms such as coupled oscillators and Analog Ising Machines naturally solve differential equations at a fraction of the energy cost of digital computation, making them attractive for low-power generative modeling, yet a fundamental mismatch exists: modern generative models assume flexible, software-defined dynamics, whereas analog hardware imposes fixed, physics-determined differential equations with limited approximation capacity. This paper introduces Analog Interaction Systems (AIS), a unified framework for hardware-implementable dynamical systems, and empirically characterizes their expressivity gap relative to neural network baselines. Two hardware-compatible mechanisms are proposed to narrow this gap — time-varying piecewise parameters and hidden physical states — and a Wasserstein GAN training procedure is developed to enable training of these models without requiring them to follow a specific trajectory. We characterize how area and power scale with connection density and precision, showing that sparse connectivity and low-bit-width quantized parameters are necessary for practical implementation, and estimate an energy cost of $23~\mu$J per generated image for the chosen architecture, representing a 2-orders-of-magnitude improvement over digital baselines. On MNIST and Fashion-MNIST, our oscillator-based \modelabbr{} achieves FID scores of 27.6 and 80.8, outperforming the best prior hardware-implementable analog generative models by $3\text{--}4\times$ with a 4-bit sparse architecture.                                                                                              
  
\end{abstract}

\maketitle 
\pagestyle{plain} 

\section{Introduction}

\begin{figure}
    \centering
    \subfloat[Conventional.]{
        \includegraphics[width=0.4\linewidth]{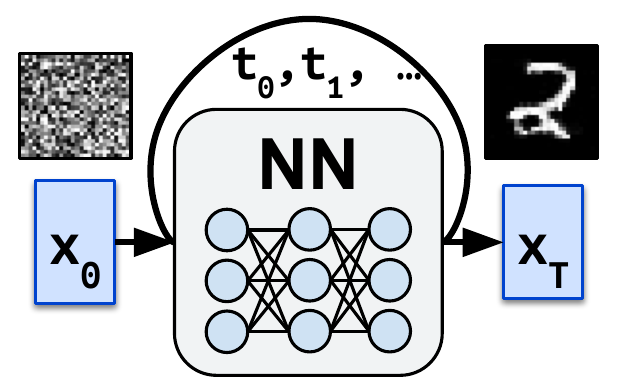}
    } 
    \subfloat[Analog Hardware.]{
        \includegraphics[width=0.6\linewidth]{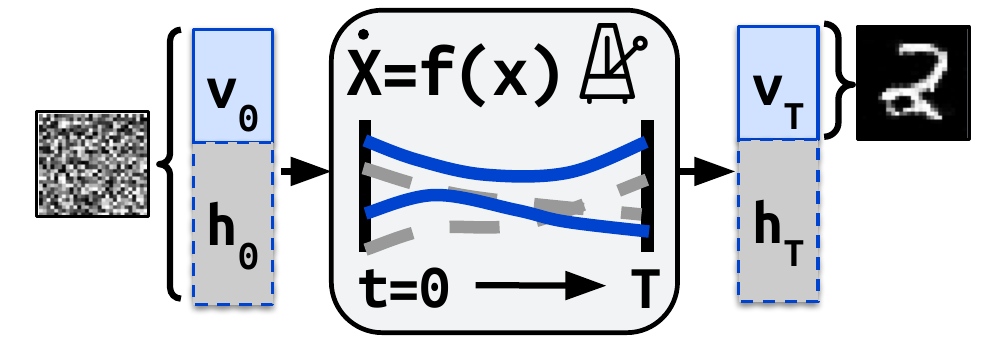}
    }
    \caption{Comparison between conventional and analog hardware image generation. The conventional method runs discrete time steps to emulate a differential equation whose vector field is computed by a neural network. The analog hardware executes a continuous differential equation whose vector field is governed by physical interactions.}
    \label{fig:overview}
\end{figure}

Analog computing offers a fundamentally different computational substrate. Rather than executing discrete arithmetic operations, analog systems allow physical state to evolve continuously under the natural laws governing their substrate—be it the phase dynamics of coupled oscillators~\cite{csaba2020osc-review}, the amplitude relaxation of optical parametric amplifiers~\cite{inagaki2016optical-ising}, or the spin settling of Ising machines~\cite{mcmahon2016coherent-ising}. Computation is performed by physics rather than by transistor switching, enabling orders-of-magnitude improvements in energy efficiency for suitable workloads. In particular, analog hardware based on interactions between physical elements, such as couplings between oscillators, has emerged as a widely studied class of dynamics-based analog hardware platforms that leverage time-evolving behaviors to solve combinatorial optimization and machine learning problems~\cite{cai2025oscnet, ahmed2021con-neighbor, mallick2021con-global, graber2024oim-local-dac, yu2025nlm,whitelam2026generative-thermodynamic-computing}. 

At the same time, modern generative models, such as diffusion models, incur substantial energy costs and have become a target workload for both researchers and companies investing in energy-efficient hardware. State-of-the-art image and audio generation systems based on diffusion and flow matching~\cite{ho2020denoising-diffusion,lipman2022flow-matching} require iterative numerical integration of learned differential equations at inference time, consuming tens to hundreds of watts per query on modern GPU hardware. Analog hardware naturally solves differential equation systems with device physics at a fraction of the energy cost of digital hardware and has recently drawn attention for low-power machine learning tasks.  Generative models are also potentially amenable to analog computation, as they are described by systems of differential equations.  \textit{Can we directly leverage the intrinsic, low-energy continuous dynamics of analog hardware to implement an energy-efficient differential equation-based generative model?}

\subsection{Challenges with Analog Generative Models}

There is a mismatch between the generative model workload and dynamics supported by analog hardware: generative models have dynamics dictated by neural networks and analog hardware implements physics-based differential equation systems based in the physical behaviors of the system (Figure~\ref{fig:overview}). These physics-based differential equation-solving hardware have several features that render them difficult to leverage for machine learning tasks:

\begin{itemize}
\item{} \textit{Non-standard Dynamics.} Because the differential equation structure is set for the physical system, it may not be good at approximating all functions. In contrast, kernels implementing neural networks are known to be good at approximating a range of functions~\cite{cybenko1989sigmoid-approximation,park1991universal-rbf, leshno1993multilayer}.
\item{} \textit{Fixed Complexity.} The complexity of the differential equation is fixed and dictated by the degree of the interactions. In contrast, the complexity of neural networks can be increased by adding layers.
\item{} \textit{Physical Limitations.} The system is subject to hardware constraints. Sparser interactions are easier to implement in hardware, parameter precision is limited, and the computation is subject to noise.
\end{itemize}

The focus of this work is as follows: \textit{Can we come up with a generative model architecture amenable to implementation on analog hardware that overcomes these limitations?}

\subsection{Analog Interaction Systems as Generative Models}

We introduce an \emph{Analog Interaction System} (\modelabbr{}), a unified model for hardware-implementable analog dynamical systems whose dynamics are fixed by device physics, and investigate their viability as generative models, developing techniques for overcoming the identified challenges. 

\prh{Expressivity.} The functional forms available to analog hardware are severely restricted by device physics. For example, oscillator networks are governed by sinusoidal coupling; Analog Ising Machines use fixed polynomial or sigmoidal nonlinearities.  We empirically characterize the expressivity gap between neural network-based dynamical systems and several \modelabbr{} model classes and identify two hardware-friendly mechanisms to close this gap: 

\begin{itemize}
\item{}\textit{Time-piecewise parameters.}  We find the complexity of \modelabbr{} behavior can be effectively increased by using time-piecewise parameters, and provide candidate analog model architectures that can implement this behavior.
\item{}\textit{Hidden physical elements.} We find the complexity of the \modelabbr{} behavior can be effectively increased by introducing hidden physical elements that interact with the physical elements used to generate the image.
\end{itemize}

We also observe that exact trajectory supervision commonly used in generative model training algorithms, such as flow matching, is infeasible for \modelabbr{} due to its limited flexibility. We instead train \modelabbr{} with a Wasserstein GAN that penalizes only the \emph{terminal} visible state, freeing the physics to discover its own path from noise to data.  This training method also supports optimizing over hidden physical elements, which lack a ground-truth trajectory to follow. On MNIST and Fashion-MNIST, our oscillator-based \modelabbr{} achieves FID scores of 27.6 and 80.8, outperforming the best prior hardware-implementable analog generative models by $3\text{--}4\times$~\cite{yu2025nlm,whitelam2026generative-thermodynamic-computing}. 

\prh{Efficiency.} We describe the hardware constraints gating the design of such systems and present a sparse \modelabbr{} analog architecture that organizes physical elements into a grid, allowing for neighboring interactions, and provide scaling laws for the power consumption of the system with increasing connectivity and parameter bit width. This architecture also has several options for implementing the time-piecewise parameters in hardware. An analog core implemented with oscillators is estimated to consume approximately $23\,\mu\text{J}$ per generated image, a 2-orders-of-magnitude improvement over digital alternatives.

\pri{Organization.} The remainder of the paper is organized as follows. Section~\ref{sec:ais} defines the analog interaction system. Section~\ref{sec:generative} formulates the generative modeling problem and presents the expressivity study, including the hidden-state and time-varying-weight experiments. Section~\ref{sec:arch} describes the hardware architecture and GAN training procedure. Section~\ref{sec:eval} evaluates generative quality against digital and analog baselines. Section~\ref{sec:related} discusses related work, and Section~\ref{sec:conclusion} concludes.



\begin{table*}[ht]
\centering
\footnotesize
\begin{tabular}{llllll}
\toprule
&Model & Dynamics & $|\theta|$ & Physical substrate & Ref. \\
\midrule
NN&MLP       & $\dot{x} = W^{(L)}\sigma(\cdots\sigma(W^{(1)}x + b^{(1)})\cdots) + b^{(L)}$ & $\sum_{i=1}^{L}(n_i^2 + n_i)$ & Standard neural ODE & --- \\
\hline
\multirow{5}{*}{\modelabbr{}}&Kuramoto  & $\dot{\theta}_i = \omega_i + \sum_j K_{ij}\sin(\theta_j - \theta_i)$ & $n+n(n-1)$ & Oscillators & \cite{csaba2020osc-review} \\
&Kuramoto+SHIL & $\dot{s}_i = {-}\lambda_{i,0}\sin(2 s_i) + \lambda_{i,1}\sin( s_i) + \sum_j C_{ij}\sin(s_i{-}s_j)$ & $2n+n \cdot (n-1)$ & Injection-locked oscillators & \cite{wang2017oim} \\
&PolyNet  & $\dot{x}_i = k_i x_i - x_i^3 + \sum_j J_{ij}x_j$ & $n+n^2$ & Analog Ising Machine (Poly) & \cite{bohm2021analog-ising} \\
&TanhNet  & $\dot{x}_i = -x_i + \tanh\!\bigl(\sum_j J_{ij}x_j\bigr)$ & $n+n^2$ & Analog Ising Machine (Tanh) & \cite{bohm2021analog-ising} \\
&SeluNet   & $\dot{x}_i = \mathrm{selu}(\sum_j W_{ij}x_j + b_i)$ & $n+n^2$ & Single-layer neural ODE & --- \\
\bottomrule
\end{tabular}
\caption{Interaction kernels for the model families studied.
  Physical models involve at most pairwise ($r=2$) state interactions;
  the MLP baseline can be deepened to increase interaction order.}
\label{tab:models}
\end{table*}

\section{Analog Interaction Systems}\label{sec:ais}

\pri{On Analog Computing.} Analog computing offers compelling advantages over digital systems for inference workloads: energy consumption scales with the physics of computation rather than with bit-precision arithmetic, enabling orders-of-magnitude improvements in energy efficiency~\cite{csaba2020osc-review}. Dedicated analog fabrics have demonstrated sub-milliwatt inference at high throughput for combinatorial optimization and matrix-vector products—workloads central to modern machine learning~\cite{inagaki2016optical-ising, mcmahon2016coherent-ising}. The state of the system is retained in properties of physical components, such as capacitors or oscillators. In analog hardware that leverages time-varying dynamics, these properties of physical components naturally evolve over time in accordance with a differential equation without requiring digital clocking or explicit MACs. This physical execution operates at a fraction of the energy and latency of digital emulation. After the system has evolved for a fixed period of time, the computed result is obtained by measuring the states of the components in the system.

\prh{Analog Interaction Systems.} We introduce the concept of an \textit{analog interaction system} (\modelabbr{}), a class of dynamical systems implementable with analog hardware that leverage interactions between devices to perform computation. In this class of systems, physical components, such as oscillators, influence one another with \textit{interactions} governed by device physics. The behavior of these systems over time can be described by an ordinary differential equation system with $n$ stateful variables $x\in\mathbb{R}^n$ that map to properties of the physical components, such as phase or amplitude:

\begin{equation}\label{eq:phys-ode}
  \dot{x}_i = L(Z^0_\theta, \sum_{j} Z^{(1)}_\theta(x_j), \sum_{j,k} Z^{(2)}_\theta(x_j, x_k), \cdots)
\end{equation}

\pri{Interactions.} Each component is subject to a number of \textit{interactions} $Z^{(p)}$ that are implemented with active components in the analog hardware substrate. The order of the interaction $r$ determines how many physical components are involved in the interaction. First-order interactions ($Z^{(1)}(x_j)$) involve a single physical component, second-order interactions ($Z^{(2)}(x_j,x_k)$) involve two physical components, and so on. For many hardware-implementable \modelabbr{} models, only first and second order interactions are supported. Each interaction is parameterized by a set of learnable parameters $\theta$, which may be programmed at runtime with digital-to-analog converters or set at fabrication time. The specific structure of the interactions is determined primarily by the physics of the system.

\pri{Laws.} The analog hardware is subject to physical laws $L(..)$ which are innately present in the physical system and naturally apply to properties of the physical components and can act on any number of elements. Examples of laws include Kirchhoff's law, which enables the summation of currents, and signal saturation, which saturates signals that fall outside a range and is implemented with a $\tanh$ function. Laws may operate on a subset of interactions.

\prh{\modelabbr{} Models.} Table~\ref{tab:models} presents the \modelabbr{} models for five kinds of interaction-based analog compute platforms:

\begin{itemize}
    \item \textit{Oscillator-based Computing.} Oscillator networks, such as CMOS ring oscillators and spin-torque nano-oscillators, solve combinatorial optimization and pattern recognition problems with low power. Oscillator-based fabrics use sinusoidal phase coupling~\cite{csaba2020osc-review}. The Kuramoto model captures standard phase coupling $K_{ij}\sin(\theta_j - \theta_i)$, while Kuramoto+SHIL (KuraSHIL) extends the model with injection-locking terms that stabilize phases to superharmonics of a reference signal, enabling binary-valued phase states suitable for Ising problems.
    \item \textit{Analog Ising Machines (AIMs).} Analog Ising Machines are physical systems engineered to minimize an Ising Hamiltonian, making them natural solvers for NP-hard combinatorial optimization. Coherent Ising Machines based on optical parametric oscillators~\cite{inagaki2016optical-ising} and degenerate optical cavities~\cite{mcmahon2016coherent-ising} have demonstrated competitive performance on MAX-CUT and graph partitioning at scale. Analog Ising machines realize this through sigmoidal or polynomial activation functions~\cite{bohm2021analog-ising}; in this work, we study both: TanhNet uses a recurrent $\tanh$ nonlinearity, and PolyNet uses a cubic soft-spin relaxation where the cubic term stabilizes amplitudes.
    \item \textit{SeluNet.} SeluNet is a single-layer neural ODE baseline, included as a non-physics-constrained reference. Unlike \modelabbr{} models, its nonlinearity is not fixed by hardware physics, providing a lower bound on the expressivity cost of the physical interaction structure.
\end{itemize}

\prh{\modelabbr{} for Machine Learning.} In this work, we focus on leveraging \modelabbr{} for machine learning tasks. In this setting, the parameters of the differential equation are learned to perform the desired computation. A key factor determining \modelabbr{}'s success as a machine learning model lies in its ability to approximate a range of functions when reparametrized. Table~\ref{tab:models} compares the structure of \modelabbr{} models to a differential equation model that uses a multi-layer perceptron (MLP) of similar interaction complexity, which is known to be a good universal approximator of functions\cite{cybenko1989sigmoid-approximation}. Observe that \modelabbr{} models use very different non-linearities than their MLP counterparts, and that these non-linearities are fixed since they arise from physical interactions. In contrast, MLPs are software-defined and can always be engineered to have higher complexity: MLPs become better approximators with more layers. \modelabbr{} models do not have an equivalent mechanism for increasing the complexity of the right-hand side of the equation.

\section{Training \modelabbr{} as Generative Models}\label{sec:generative}

Generative flow matching casts learning as a problem of transporting samples from a simple prior $p_0$ to a data distribution $p_1$ via a differential equation~\cite{lipman2022flow-matching, tong2023improving}:
\begin{equation}
    \frac{d\vecstatevar}{dt} = f_\theta(\vecstatevar, t),
\end{equation}
where $f_\theta$ is a parameterized vector field. An \modelabbr{} model serves naturally as such a transport map: given an initial visible state $v(0) \sim p_0$, the system integrates forward under its learned dynamics, and the terminal state $v(T)$ is read out as the generated sample. Unlike digital flow matching—where the vector field is re-evaluated at every solver step—the analog system requires only a single physical settling period of duration $T$, making inference energy cost independent of the number of integration steps. However, \modelabbr{}'s interaction kernels are fixed by hardware physics, and this structural constraint limits the set of trajectories the model can express. Three mechanisms recover expressivity: (1)~the training objective, (2)~time-piecewise weights, and (3)~hidden state augmentation. We study these mechanisms on low-dimensional synthetic distribution datasets.

\begin{figure}[t]
\centering
\subfloat{%
  \includegraphics[width=\linewidth]{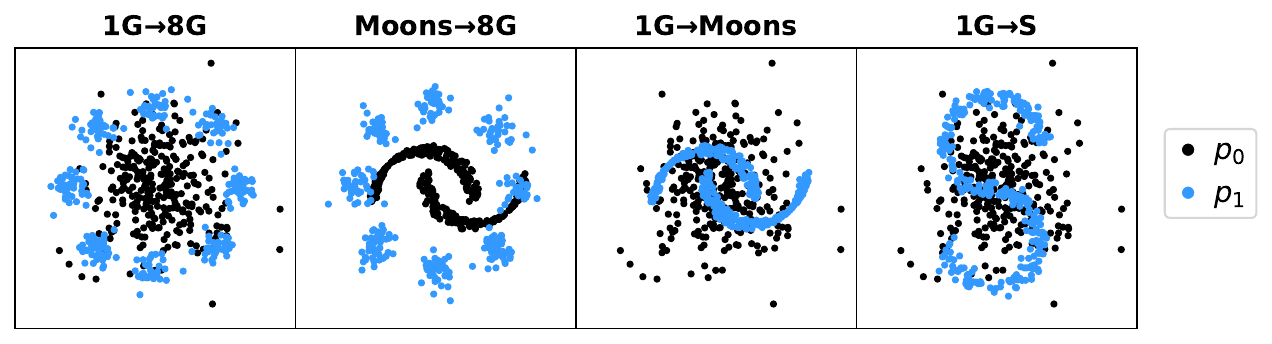}%
}\\[4pt]
\subfloat{%
  \resizebox{\linewidth}{!}{%
  \begin{tabular}{lll}
  \toprule
   & Source $p_0$ & Target $p_1$ \\
  \midrule
  $1\text{G}{\to}8\text{G}$       & Gaussian  & 8-component Gaussian mixture \\
  $\text{Moons}{\to}8\text{G}$   & Two moons & 8-component Gaussian mixture \\
  $1\text{G}{\to}\text{Moons}$   & Gaussian  & Two moons \\
  $1\text{G}{\to}S$              & Gaussian  & S-curve manifold \\
  \bottomrule
  \end{tabular}}%
}
\caption{Distribution pairs used for generative fitting experiments.
  Top: source $p_0$ (black) and target $p_1$ (blue) scatter for each pair.
  Bottom: summary table.}
\label{fig:pairs}
\end{figure}

\begin{figure}[t]
\centering
\includegraphics[width=\linewidth]{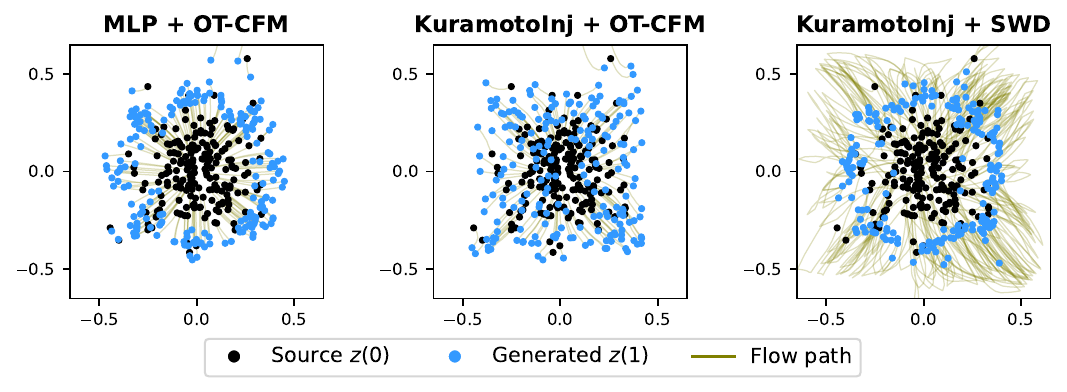}
\caption{Distributional fitting on $1\text{G}{\to}8\text{G}$.
  From left: MLP + OT-CFM, KuraSHIL + OT-CFM, KuraSHIL + SWD.
  Unlike the flexible MLP, the physics-constrained \modelabbr{} cannot follow
  straight-line OT-CFM paths; SWD decouples trajectory shape from target
  quality and recovers transport.}
\label{fig:flows_insight1}
\end{figure}

\subsection{Experimental Setup}\label{sec:gen:setup}

\prh{Distribution Pairs.}
The four transport tasks (Figure~\ref{fig:pairs}) cover unimodal, multimodal, and manifold target distributions. Data are normalized to $[-0.5, 0.5]^2$.

\prh{Models and Conditions.}
Each \modelabbr{} model is evaluated in two configurations. (a) No hidden state: the model operates on the 2-dimensional visible state alone, with the integration interval $[0,1]$ divided into $c \in \{1, 2, 4\}$ equal segments, each assigned an independent parameter copy. (b) With hidden state: the state is augmented to $z = [v;\,h] \in \mathbb{R}^{2+k}$, $k \in \{2, 4\}$, always with $c = 4$ segments; the latent initial condition $h_0 \sim \mathcal{N}(0,\, 0.25\,I)$ is sampled independently per forward pass, and only the visible dimensions $v(T)$ are read out as the generated sample. Under OT-CFM, $h_0$ is additionally conditioned on the coupled source--target pair $(x_0, x_1)$; under SWD, the stochastic $h_0$ alone is used since no paired supervision is available.

\prh{Training Objective Functions.}
Modern generative models achieve state-of-the-art quality by training a time-conditioned vector field $f_\theta(\vecstatevar, t)$ to transport a prior $p_0$ to a target $p_1$. Diffusion models~\cite{ho2020denoising-diffusion, song2020score-based-diffusion} parameterize this as a stochastic reversal of a fixed noising process; flow matching methods~\cite{lipman2022flow-matching} construct deterministic transport paths between coupled samples. Both impose \emph{trajectory supervision}: the model must match a prescribed velocity field at every intermediate time $t$.

For example, optimal-transport conditional flow matching (OT-CFM), a state-of-the-art trajectory supervision algorithm for generative modeling, constructs a minibatch OT coupling between source and target samples $(x_0, x_1)$ at each step, and regresses the model velocity onto the straight-line target $u_t = x_1 - x_0$ at interpolated states $x_t = (1-t)x_0 + t x_1$~\cite{tong2023improving}:
\begin{equation}
\mathcal{L}_\text{CFM}(\theta) = \mathbb{E}_{t,x_0,x_1}\bigl[\|f_\theta(x_t,t) - (x_1-x_0)\|^2\bigr].
\end{equation}
This loss constrains the model to straight paths, regardless of what trajectories the \modelabbr{} physics would naturally produce.

\emph{Endpoint supervision} instead constrains only the terminal distribution $v(T) \sim p_1$, leaving the \modelabbr{} physics free to traverse any path from noise to data.
We realize this via Sliced Wasserstein Distance (SWD). The Wasserstein distance quantifies the minimum transport cost between two distributions—the minimum average displacement required to rearrange one into the other. The sliced variant approximates this by projecting samples onto random unit vectors and averaging the resulting 1D Wasserstein distances:
\begin{equation}
\mathcal{L}_\text{SWD}(\theta) = \mathbb{E}_{\omega \sim \mathcal{U}(\mathbb{S}^{d-1})}\bigl[W_1(\omega^\top v(T),\, \omega^\top x_1)\bigr].
\end{equation}
SWD imposes no constraint on intermediate trajectories, allowing the \modelabbr{} physics to self-organize over the full integration interval rather than being forced to produce straight-line paths at each instant.

\pri{Hyperparameter Tuning and Evaluation.}
Each model runs an independent Optuna study with 30 trials of 4\,000 training steps each to tune the learning rate and magnitude of the randomly initialized weights. The dynamical system is solved with standard Runge--Kutta solver and evaluated with the squared 2-Wasserstein distance computed via exact optimal transport:
\begin{equation}
W_2^2(p, q) = \inf_{\gamma \in \Gamma(p,q)} \mathbb{E}_{(x,y)\sim\gamma}\bigl[\|x-y\|^2\bigr],
\end{equation}
where $\Gamma(p,q)$ is the set of couplings with marginals $p$ and $q$. We estimate $W_2^2$ from 256 samples over 3 random seeds. Lower $W_2^2$ indicates better transport quality.

\subsection{Distributional Fitting Results}\label{sec:gen:results}
\begin{figure*}[htbp]
\centering
\includegraphics[width=\textwidth]{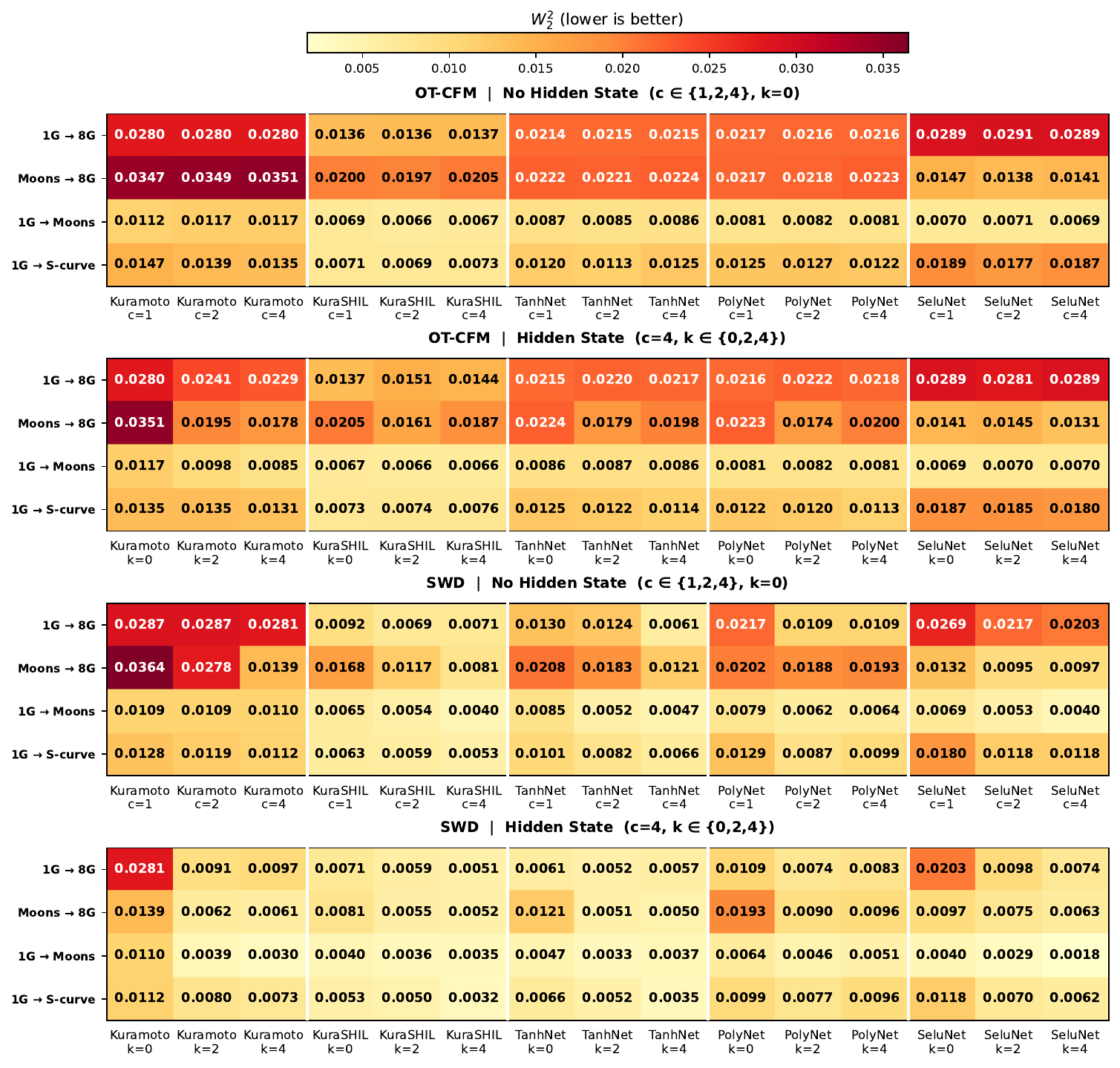}
\caption{$W_2^2$ results across all conditions, model classes, and distribution pairs (lower is better). Each panel shows one training condition; columns group model classes by \modelabbr{} model ($c$ = chunk count, $k$ = hidden dimensions). }
\label{fig:w2_heatmap}
\end{figure*}

Figure~\ref{fig:w2_heatmap} shows $W_2^2$ scores on the distributional fitting tasks. Each panel is a heatmap: rows are distribution pairs, columns are model variants, and cell shade encodes $W_2^2$, with lighter cells indicating lower values (better fit). Panels are organized by training objective (OT-CFM top, SWD bottom) and hidden-state configuration (w/ and w/o hidden state).

\prh{Improvement 1: Endpoint Supervision.}
SWD consistently outperforms OT-CFM across all distribution pairs and model classes. Figure~\ref{fig:flows_insight1} illustrates why: under OT-CFM, the physics-constrained AIS produces scattered trajectories that fail to reach the target shape, while SWD allows the dynamics to self-organize into structured transport. The gap is particularly pronounced on multimodal targets (D1, D2): OT-CFM forces near-straight-line paths that conflict with the constrained interaction structure, while SWD decouples trajectory shape from target quality. This result indicates that \emph{endpoint supervision is the appropriate objective for \modelabbr{}-based generative models}.

\prh{Improvement 2: Time-Piecewise Weights.}
A model with fixed parameters must express the entire source-to-target transport with a single vector field parameterization.
Under OT-CFM the benefit is marginal: the straight-line velocity target already provides strong directional supervision, leaving little room for chunk-wise specialization. Under SWD the gains are clearer. For example, KuraSHIL on $\text{Moons}{\to}8\text{G}$ improves from $W_2^2 = 0.0168$ at $c{=}1$ to $W_2^2 = 0.0081$ at $c{=}4$ because each chunk can adapt its vector field to a different phase of the transport without conflicting with an intermediate-step constraint.

\prh{Improvement 3: Hidden States.}
Adding $k$ hidden dimensions to the visible state directly expands the model's effective parameter space: the learned dynamics operate in $\mathbb{R}^{2+k}$ rather than $\mathbb{R}^2$, increasing both the number of learnable weights and the dimensionality of the vector field.
Hidden-state SWD consistently yields the lowest $W_2^2$ values. For example, on $1\text{G}{\to}\text{Moons}$ with $k{=}4$ hidden dimensions, $W_2^2$ drops from $0.0040$ to $0.0018$. Hidden state increases the model's expressivity, while freeing the objective from trajectory constraints allows that expressivity to be exploited.

\prh{\modelabbr{} Model Comparison.} Not all \modelabbr{} models perform equally. 
Among the five model classes, KuraSHIL is the strongest performer under both objectives, achieving the best or near-best $W_2^2$ on $1\text{G}{\to}8\text{G}$, $1\text{G}{\to}\text{Moons}$, and $1\text{G}{\to}S$ across conditions. Its per-node injection-locking terms provide additional degrees of freedom beyond the Kuramoto model, enabling finer-grained control of individual oscillator phases. Kuramoto is the weakest on multimodal targets $1\text{G}{\to}8\text{G}$ and $\text{Moons}{\to}8\text{G}$. SeluNet performs well on $1\text{G}{\to}\text{Moons}$ and $1\text{G}{\to}S$ but poorly on $1\text{G}{\to}8\text{G}$. TanhNet and PolyNet occupy a consistent middle ground, achieving moderate $W_2^2$ across all pairs.

\subsection{Training with Generative Adversarial Networks} 
The distributional fitting experiments above confirm endpoint supervision as the right training paradigm for \modelabbr{}.
In high dimensions, however, directly estimating the Wasserstein distance is computationally intractable, as exact optimal transport scales poorly with dimensionality.
The Wasserstein Generative Adversarial Networks (WGAN) formulation~\cite{gulrajani2017wgan-gp} addresses this by learning a discriminator to compute approximated Wasserstein distance.
The discriminator evaluates only the generated output (terminal state $v(T)$), never intermediate trajectories, making the WGAN objective an endpoint supervision method by design.

A standard GAN pits two neural networks working against each other: a generator that attempts to transform random noise into realistic data, and a discriminator that attempts to distinguish between the generated fakes and real data samples. Through this competitive process, the generator iteratively learns to approximate the true data distribution. In our framework, the core idea is simple: we replace the neural network generator with the  Analog Interaction System. To compute the "forward pass" during training, we run a differentiable dynamical system simulation (an ODE solver) from $t=0$ to $t=T$. By differentiating through this simulation, we can backpropagate the discriminator's feedback directly into the physical coupling weights of the \modelabbr{}. 

We utilize the Wasserstein GAN formulation with Gradient Penalty (WGAN-GP) as detailed in Algorithm~\ref{alg:wgan_agm}~\cite{gulrajani2017wgan-gp} .
$D_w$ is the digital discriminator network, and $\text{Proj}_{vis}$ extracts the visible node states from the full analog system state $x(T)$. The training loop alternates between updating the digital discriminator and the analog generator parameters. First, the discriminator $w$ is trained for $n_{\text{disc}}$ steps to maximize the Wasserstein distance between the real data batch and the generated batch $\hat{x}$, regularized by a gradient penalty. Subsequently, the generator step updates the physical analog parameters $\theta$. We sample initial noise, simulate the analog dynamics forward to time $T$, and evaluate the generated samples using the discriminator. We then compute the loss gradients with respect to $\theta$ by backpropagating through the ODE solver, pulling the analog dynamics' terminal states closer to the true data manifold.

\begin{algorithm}[tb]
\caption{WGAN-GP Training of \modelabbr{}.} 
\small
\label{alg:wgan_agm}
\textbf{Require:} Batch size $m$, learning rate $\alpha$, discriminator iterations $n_{\text{disc}}$, initial discriminator parameters $w$, initial analog parameters $\theta$, integration time $T$, analog dynamics function $f_\theta$

\begin{algorithmic}[1]
\While{$\theta$ has not converged}
    \For{$t = 1, \dots, n_{\text{disc}}$}
        \For{$i = 1, \dots, m$}
            \State Sample real data $\{x^{(i)}\}_{i=1}^m \sim \mathbb{P}_{data}$
            \State Sample initial states $\{x_0^{(i)}\}_{i=1}^m \sim \mathbb{P}_{prior}$
        
            \State $\hat{x}^{(i)} \leftarrow \text{ODESolve}(x_0^{(i)}, f_\theta, T)$
            \State $L_w \leftarrow D_w(\hat{x}^{(i)}) - D_w(\text{Proj}_{vis}(x^{(i)})) +$
            \State $\lambda \cdot\text{GradientPenalty}(D_w, x, \hat{x})$
        \EndFor
        \State $w \leftarrow \text{Adam}(\alpha, L_w)$
    \EndFor
    
    \State Sample initial states $\{x_0^{(i)}\}_{i=1}^m \sim \mathbb{P}_{prior}$
    \State $L_\theta \leftarrow -\frac{1}{m} \sum_{i=1}^m D_w(\text{Proj}_{vis}(\text{ODESolve}(x_0^{(i)}, f_\theta, T)))$
    \State $\theta \leftarrow \text{Adam}(\alpha, L_\theta)$
\EndWhile
\end{algorithmic}
\end{algorithm}

\section{Hardware-Efficient \modelabbr{} Architecture}\label{sec:arch}

While analog computing paradigms promise benefits in energy efficiency and computational latency, the realization of such systems is fundamentally governed by strict physical constraints. 
In this section, we first analyze the hardware constraints of analog dynamical systems, including interaction complexity, routing capacity, parameter programmability, and noise. We then present our hardware-efficient \modelabbr{} architecture that operates within these constraints. Finally, we estimate power consumption and derive scaling laws for connectivity and weight precision based on published silicon measurements.

\subsection{Analysis of Hardware Constraints}
The dynamics of the original \modelabbr{} (Equation~\ref{eq:phys-ode}) with hardware constraints included can be stated as follows:
\begin{equation}\label{eq:hw-model}
\dot{x}_i = L(Z^{(0)}_{\hat\theta(t)}(x_i) , \sum_{j\in \mathcal{N}_i} Z^{(1)}_{\hat\theta(t)}(x_j), \sum_{j\in \mathcal{N}_i} Z^{(2)}_{\hat\theta(t)}(x_i, x_j), \noiseamp\mathcal{N}(0, 1))
\end{equation}
\prh{Interaction Complexity ($Z^{(0)}$, $Z^{(1)}$, $Z^{(2)}$).}
The model restricts state interactions to at most second-order terms. Signals interact via the physical summation of these low-order terms, modeling hardware-native aggregation mechanisms such as Kirchhoff's current law on a common wire. Arbitrarily high-order multivariate interactions are physically prohibitive and therefore excluded.

\prh{Routing Capacity ($\mathcal{N}_i$).}
Dense routing incurs high area overheads. For example, in oscillator-based Ising machines utilizing a King's graph topology, the active oscillators occupy less than 5\% of the total chip area, with the vast majority consumed by coupling and routing circuitry, even though the connectivity is already sparse~\cite{moy2022-1968node-con}. Increasing the connection density worsens the situation. Comparing a fully-connected network to a nearest neighbor network on the same $65$ nm CMOS technology and area $1.44 \text{ mm}^2$, the former supports only 30 oscillators, whereas the latter contains 560 oscillators~\cite{mallick2021con-global, ahmed2021con-neighbor}. 
Consequently, the state $x_i$ connects only to a restricted subset of neighbors $\mathcal{N}_i$, comprising local connections and a limited number of long-distance routes.

\prh{Parameter Programmability ($\hat\theta(t)$).}
High-resolution programmability requires digital-to-analog converters (DACs) that consume significant die area and power. Therefore, parameters are restricted to low resolution in both bit-width and time. 
We model a parameter $\hat\theta$ as a function over $\temporalweight$ time chunks with quantization:
\begin{equation}
\hat\theta(t) = \sum_{k=1}^{\temporalweight} \theta^k \cdot \mathbb{I}_{[t_k, t_{k+1})}(t), \quad \theta^k=\mathcal{Q}_\weightbit(w_k)
\end{equation}
where $w_k \in \mathbb{R}$ is the target weight for the $k$-th interval, $\mathcal{Q}_\weightbit(\cdot)$ is a $\weightbit$-bit quantization operator, and $\mathbb{I}_{[t_k, t_{k+1})}(t)$ is an indicator function that evaluates to $1$ when $t \in [t_k, t_{k+1})$ and $0$ otherwise.

\prh{Noise ($\noiseamp\mathcal{N}(0, 1)$).} 
Analog hardware intrinsically exhibits noise, which introduces random and small-magnitude perturbations throughout the system. We incorporate an additive Gaussian noise $\noiseamp\mathcal{N}(0, 1)$ into the dynamics to capture this physical phenomenon.

\subsection{Proposed Architecture}

\begin{figure}[t]
    \centering
    
    \begin{minipage}{0.55\linewidth}
        \centering
        \subfloat[AIS Architecture]{
            \includegraphics[width=0.75\linewidth]{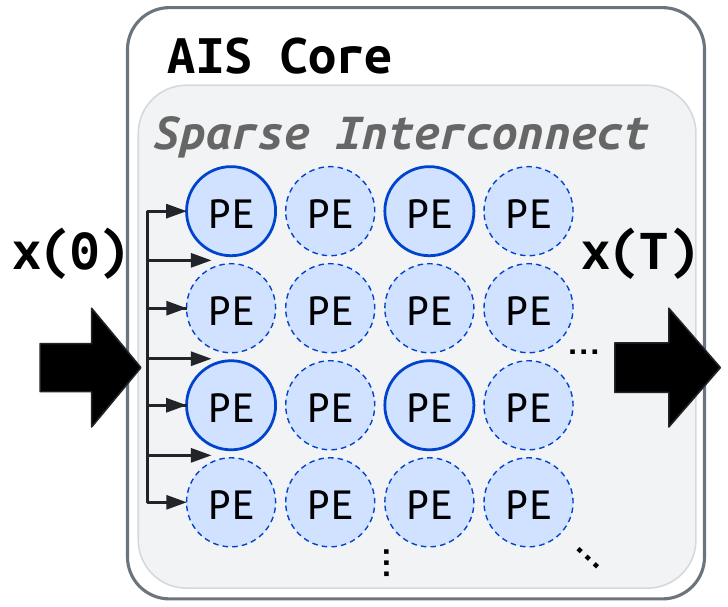}
            \label{fig:ais_arch}
        }
    \end{minipage} 
    \begin{minipage}{0.3\linewidth}
        \centering
        \subfloat[Oscillator-based PE]{
            \includegraphics[width=\linewidth]{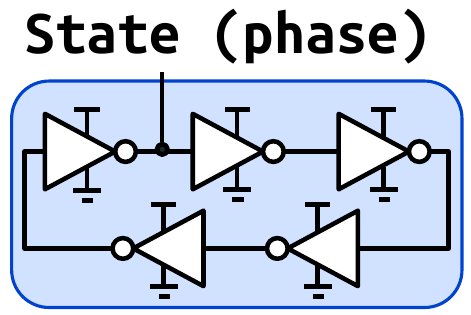}
            \label{fig:ais_osc}
        }
        
        \vspace{-1em}
        
        \subfloat[Integrator-based PE]{
            \includegraphics[width=\linewidth]{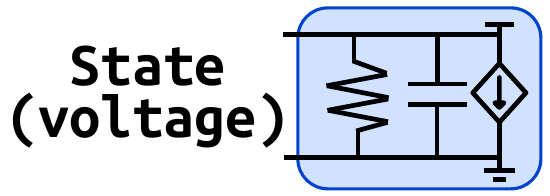}
            \label{fig:ais_int}
        }
    \end{minipage}

    \subfloat[Asymmetric Coupling with Inverter Arrays.]{
        \includegraphics[width=0.65\linewidth]{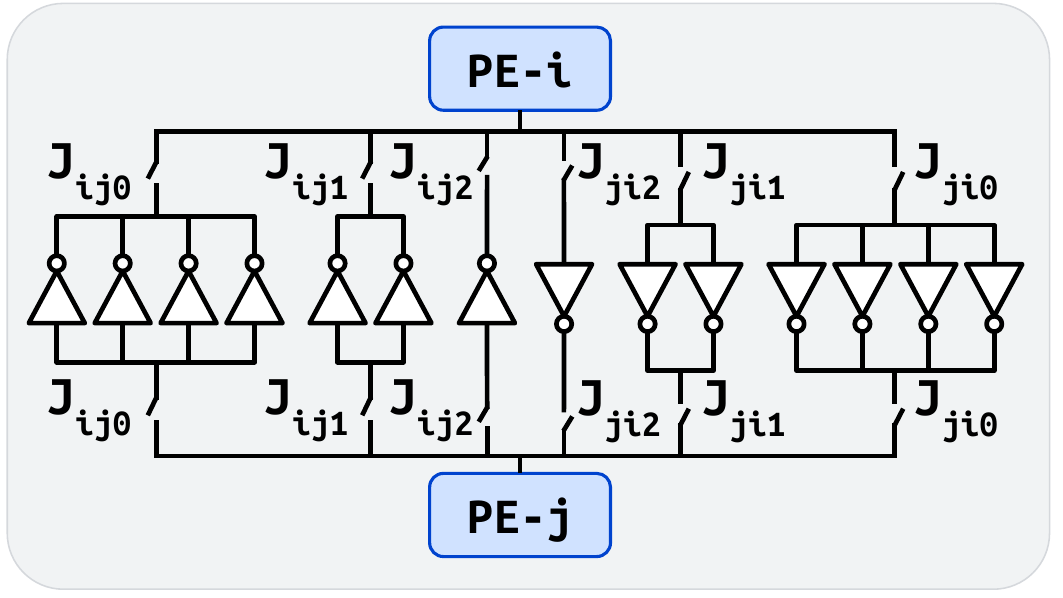}
        \label{fig:inv_cpl}
    }
    \caption{Overview of the \modelabbr{} Architecture.}
    \label{fig:agm-topology}
\end{figure}

Our approach is conceptually straightforward: design a physical architecture to operate within the aforementioned constraints while maintaining the capacity for generative modeling. The core of an \modelabbr{} machine contains two types of components: compute nodes (Physical Elements, PE) and programmable coupling units.

\prh{Topology and State Partitioning.}
We map the system states onto a two-dimensional grid of physical elements, forming the \modelabbr{} core. The physical elements are partitioned into visible elements, which represent the target data dimensions, and hidden elements.
For example, in Figure~\ref{fig:ais_arch}, the grid consists of repeating structural tiles where visible nodes (solid circles) are surrounded by hidden nodes (dotted circles). Programmable coupling units then wire these elements together according to a predefined neighborhood, and this is repeated for every PE.

\prh{Physical Elements.} 
The PEs store the state variables and evolve them natively according to their intrinsic circuit response or device physics, as formalized in Equation~\ref{eq:hw-model}. From an architectural perspective, a PE can be implemented using any component that exhibits dynamics that designers find suitable. For example, coupled oscillators naturally realize Kuramoto dynamics (Figure~\ref{fig:ais_osc}), and analog integrators with saturation non-linearities can implement the AIM-Tanh dynamics (Figure~\ref{fig:ais_int}).

\begin{figure}[t]
    \centering
    
    \subfloat[Save-and-Reprogram]{
        \includegraphics[height=60pt]{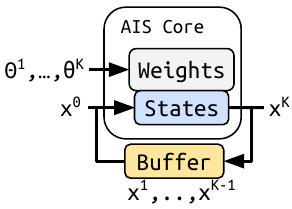}
        \label{fig:save_reprog}
    }
    \subfloat[State Stationary]{
        \includegraphics[height=60pt]{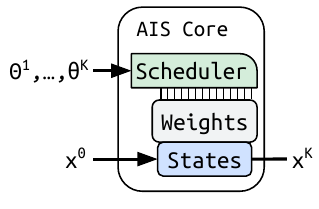}
        \label{fig:state_stat}
    }

    \subfloat[Weight Stationary]{
        \includegraphics[height=60pt]{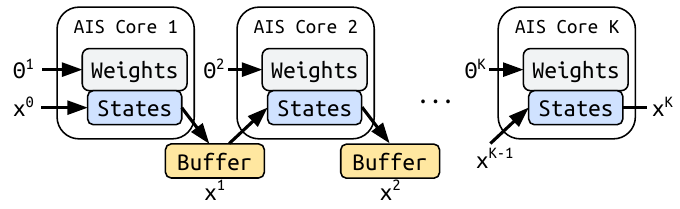}
        \label{fig:weight_stat}
    }
    
    \caption{Architectural strategies for time-varying weights.}
    \label{fig:time_varying_strategies}
\end{figure}

\prh{Programmable Coupling Units.}
The coupling units transmit interacting state information between PEs, scaled by programmable weights. Asymmetric coupling ($J_{ij} \neq J_{ji}$) is realized by deploying two independent, unidirectional coupling elements between interacting PEs, with each direction configured by a dedicated DAC. For example, in an oscillator-based \modelabbr{} implementation, this unidirectional interaction is realized using programmable inverter arrays (Figure~\ref{fig:inv_cpl}). 
The $\temporalweight$-time-varying coupling can be implemented with three strategies, each with hardware trade-offs (Figure~\ref{fig:time_varying_strategies}):
\begin{itemize}
    \item \textit{Save-and-reprogram}: At the end of each temporal interval, the intermediate state variables are read out, and the weights are reprogrammed with the next set of values. This approach requires only a single \modelabbr{} core but incurs latency and energy overheads due to the repeated digitization and writing of intermediate analog states.
    \item \textit{Weight stationary}: This approach utilizes spatial unrolling, mapping the $\temporalweight$ discrete time steps across $\temporalweight$ statically configured \modelabbr{} cores. The state variables are read out and pipelined sequentially from one core to the next. This eliminates the need to dynamically reprogram the DACs but necessitates inter-core state read/write operations and requires $\temporalweight$ times more physical hardware.
    \item \textit{State stationary}: The analog state variables remain within a single \modelabbr{} core, evolving continuously without interruption. The local coupling units are internally scheduled to update their weights dynamically at predefined temporal boundaries. This approach avoids intermediate state digitization but requires integrated memory with timing control circuitry at each coupling unit to manage the synchronous weight updates.
\end{itemize}

\subsection{Power Estimation}
To establish a coarse-grained power estimate, we extrapolate from the 28-nm coupled oscillator Ising machine presented by Graber and Hofmann~\cite{graber2024oim-local-dac}.
Their silicon implementation features 1,440 nodes with an average degree of 8 couplings per node using 4-bit resolution DACs, consuming $460.3~\text{mW}$ of continuous power, giving a per-node baseline of:
\begin{equation}
    P_{\text{node,ref}} = \frac{460.3~\text{mW}}{1440~\text{nodes}} \approx 320~\mu\text{W}/\text{node}
\end{equation}
The chip's component breakdown further reports $113.3~\mu\text{W}$ per oscillator periphery and ${\approx}23~\mu\text{W}$ per coupler; these align at the 8-coupling degree ($113.3 + 23 \times 8 \approx 297~\mu\text{W}$), with the ${\approx}7\%$ gap likely attributable to global routing overhead.

For an \modelabbr{} core with $M$ nodes and $L$ couplings per node, we extrapolate by scaling the coupler count and applying a $1.1\times$ power penalty for global routing:
\begin{equation}
    P_{\text{node}}(L) \approx 113.3 + 23 \times L \times 1.1 \ \mu\text{W}
\end{equation}

For a $56\times56$ \modelabbr{} core generating MNIST images with $L=24$ couplings per node:
\begin{align}
    P_{\text{node}}(24) &\approx 113.3 + 23 \times 24 \times 1.1 \approx 721~\mu\text{W} \\
    P_{\text{total}} &= 56 \times 56 \times 721~\mu\text{W} \approx 2.3~\text{W}
\end{align}
Operating at $100~\text{MHz}$ with $1000$ cycles per sample, the energy per generated image is:
\begin{equation}
    E_{\text{image}} \approx \frac{2.3~\text{W} \times 1000~\text{cycles}}{100~\text{MHz}} = 23~\mu\text{J}
\end{equation}
This is 2 orders of magnitude more energy efficient than 7--79~$\text{mJ}$ per image estimated for digital generative models~\cite{chen2025optical-generative-model}\footnote{Precise power consumption depends heavily on layout considerations, e.g., routing area growth, which are not modeled here and may increase power.}.

To generalize these estimates, we parameterize an \modelabbr{} architecture by node count $M$, coupling degree $L$, and weight bit-width $q$. In programmable-inverter-based coupling, each additional bit doubles the number of required drive elements; we capture this as a $2^{q-4}$ factor relative to the 4-bit reference:
\begin{equation}
    E_{\text{image}}(M, L, q) \approx M \cdot (113.3 + 23 \times L \times 1.1) \cdot 2^{q-4} \times 10^{-5}~\mu\text{J}
\end{equation}
Switching to all-to-all connectivity ($L = M-1 = 3135$) raises $P_{\text{node}}$ by ${\approx}110\times$, yielding $E \approx 2.5~\text{mJ}$; because energy scales linearly with $L$, all-to-all connectivity causes total energy to grow quadratically with $M$, becoming physically prohibitive at scale. Increasing weight resolution to $q{=}8$ bits multiplies energy by $2^4 = 16\times$, giving $E \approx 0.36~\text{mJ}$.
These trends confirm that sparse connectivity and low bit-width are fundamental requirements for energy-efficient analog generative modeling.

\section{Evaluation}\label{sec:eval}

\begin{table}[htbp]
\centering
\small
\caption{Resource usage for the studied generative models. Cpl/Asym/Sym abbreviate coupling/asymmetric/symmetric, respectively. $\temporalweight$ is the number of chunks of time-varying weights.}
\label{tab:resource_config}
\begin{tabular}{@{}lcccc@{}}
\toprule
Arch & \# PE & Cpl / PE  & \# Param & \# Cpl \\ 
\midrule
\multicolumn{5}{@{}l}{\textit{\modelabbr{} (Ours)}} \\
\hspace{2pt} Asym., $\temporalweight=1$    & $56\times56$  & 24  & 81536   & 75264 \\
\hspace{2pt} Asym., $\temporalweight=2$    & $56\times56$  & 24  & 163072  & 75264 \\
\hspace{2pt} Sym., $\temporalweight=4$     & $56\times56$  & 24  & 175616  & 37632 \\
\hspace{2pt} Asym., $\temporalweight=4$    & $56\times56$  & 24  & 326144  & 75264 \\
\midrule
\multicolumn{5}{@{}l}{\textit{Baselines}} \\
\hspace{2pt} DTM   & $80\times80\times4$ & 24   & 343008  & 1718784 \\
\hspace{2pt} NLM        & $28\times28$   & 784  & 614656  & 614656 \\
\bottomrule
\end{tabular}
\end{table}

\begin{table}[t]
\centering
\small
\caption{FID scores of the studied models. $\temporalweight$: Number of chunks in time-varying weights. $\weightbit$: Weight quantization bits. $\noiseamp$: Noise strength.}
\label{tab:dse}
\begin{tabular}{lcccccc} 
\toprule
\multirow{2}{*}{Model} & \multirow{2}{*}{Asym.?} & \multirow{2}{*}{$\temporalweight$} &  \multirow{2}{*}{$\weightbit$} & \multirow{2}{*}{$\noiseamp$} & \multicolumn{2}{c}{FID $\downarrow$} \\
\cmidrule(lr){6-7}
 & & & &  & MNIST & FMNIST \\
\midrule
\multicolumn{7}{l}{\textit{\modelabbr{} (Ours)}} \\

\hline
\multicolumn{7}{l}{\footnotesize\textit{Impact of the analog model}} \\
\rowcolor{gray!15} \hspace{2pt} KuraSHIL & & & & & & \\
\rowcolor{gray!15} \hspace{8pt} \textit{(default)} & \multirow{-2}{*}{$\checkmark$} & \multirow{-2}{*}{4} & \multirow{-2}{*}{64} & \multirow{-2}{*}{0} & \multirow{-2}{*}{27.6} & \multirow{-2}{*}{80.8} \\
\hspace{2pt} Kuramoto  & $\checkmark$ & 4    & 64 & 0     & 288.77 & 144.82 \\
\hspace{2pt} PolyNet  & $\checkmark$     & 4    & 64 & 0     & 307.3 & 160.3 \\
\hspace{2pt} TanhNet  & $\checkmark$     & 4    & 64 & 0     & 90.7  & 95.4  \\
\hspace{2pt} SeluNet  & $\checkmark$     & 4    & 64 & 0     & 335.76  & 237.11 \\
\hline
\multicolumn{7}{l}{\footnotesize\textit{Impact of temporal and asymmetric weights}} \\
\hspace{2pt} KuraSHIL  & $\checkmark$     & 1    & 64 & 0     & 62.7  & 84.4  \\
\hspace{2pt} KuraSHIL  & $\checkmark$     & 2    & 64 & 0     & 42.5  & 82.6  \\
\rowcolor{gray!15} \hspace{2pt} \textit{KuraSHIL} & $\checkmark$ & 4 & 64 & 0 & 27.6 & 80.8 \\
\hspace{2pt} KuraSHIL  & $\times$         & 4    & 64 & 0     & 38.4  & 81.3 \\

\hline
\multicolumn{7}{l}{\footnotesize\textit{Impact of hidden states}} \\
\rowcolor{gray!15} \hspace{2pt}\textit{KuraSHIL} & $\checkmark$ & 4 & 64 & 0 & 27.6 & 80.8 \\
\hspace{2pt} KuraSHIL  & \multirow{2}{*}{$\checkmark$} & \multirow{2}{*}{4} & \multirow{2}{*}{64} & \multirow{2}{*}{0} & \multirow{2}{*}{43.0} & \multirow{2}{*}{84.3} \\
\hspace{2pt} (no hidden states) & & & & & & \\

\hline
\multicolumn{7}{l}{\footnotesize\textit{Impact of Quantization}} \\
\rowcolor{gray!15} \hspace{2pt}  \textit{KuraSHIL} & $\checkmark$ & 4 & 64 & 0 & 27.6 & 80.8 \\
\hspace{2pt} KuraSHIL  & $\checkmark$     & 4    & 4  & 0     & 27.2  & 81.9  \\
\hspace{2pt} KuraSHIL  & $\checkmark$     & 4    & 3  & 0     & 32.9  & 85.6  \\
\hspace{2pt} KuraSHIL  & $\checkmark$     & 4    & 2  & 0     & 53.9  & 98.5  \\
\hline
\multicolumn{7}{l}{\footnotesize\textit{Impact of Noise}} \\
\rowcolor{gray!15} \hspace{2pt}  \textit{KuraSHIL} & $\checkmark$ & 4 & 64 & 0 & 27.6 & 80.8 \\
\hspace{2pt} KuraSHIL  & $\checkmark$     & 4    & 4  & 0.025 & 29.1  & 75.1  \\
\hspace{2pt} KuraSHIL  & $\checkmark$     & 4    & 4  & 0.05  & 36.7  & 78.4  \\
\hspace{2pt} KuraSHIL  & $\checkmark$     & 4    & 4  & 0.1   & 54.5  & 90.4  \\
\midrule
\multicolumn{7}{l}{\textit{Baselines}} \\
\multicolumn{5}{c}{DTM} & 107.8 & 112.8 \\
\multicolumn{5}{c}{NLM} & 230.5 & 200.8 \\
\bottomrule
\end{tabular}
\end{table}

\begin{figure*}[tb]
    \centering
    
    \makebox[0.12\textwidth][c]{\textbf{KuraSHIL}} 
    \makebox[0.12\textwidth][c]{\textbf{KS-Nonideal}} 
    \makebox[0.12\textwidth][c]{\textbf{Kuramoto}} 
    \makebox[0.12\textwidth][c]{\textbf{TanhNet}} 
    \makebox[0.12\textwidth][c]{\textbf{PolyNet}}
    \makebox[0.12\textwidth][c]{\textbf{SeluNet}}
    \makebox[0.12\textwidth][c]{\textbf{DTM}}
    \makebox[0.12\textwidth][c]{\textbf{NLM}} \\ 
    
    \includegraphics[width=0.12\textwidth]{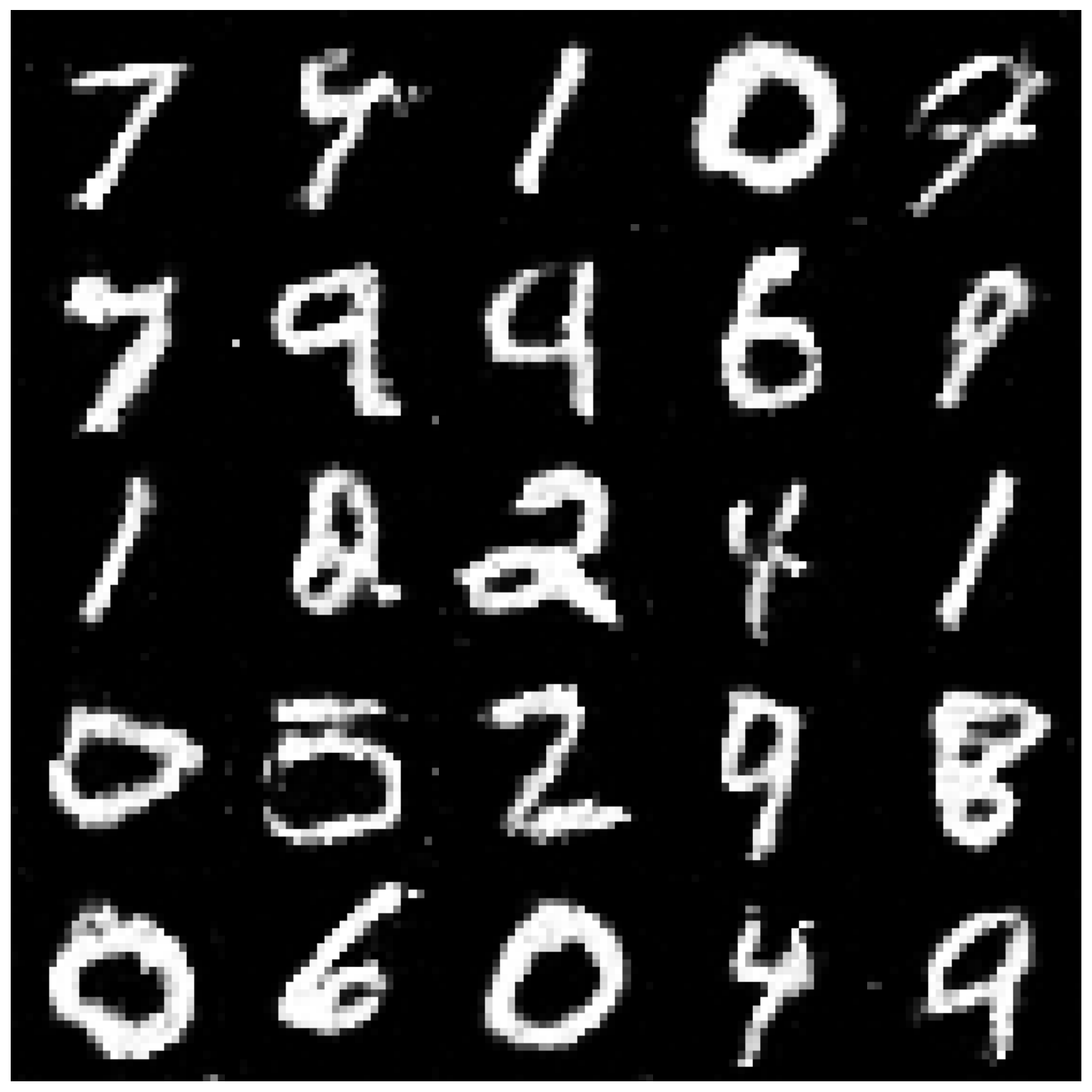}
    \includegraphics[width=0.12\textwidth]{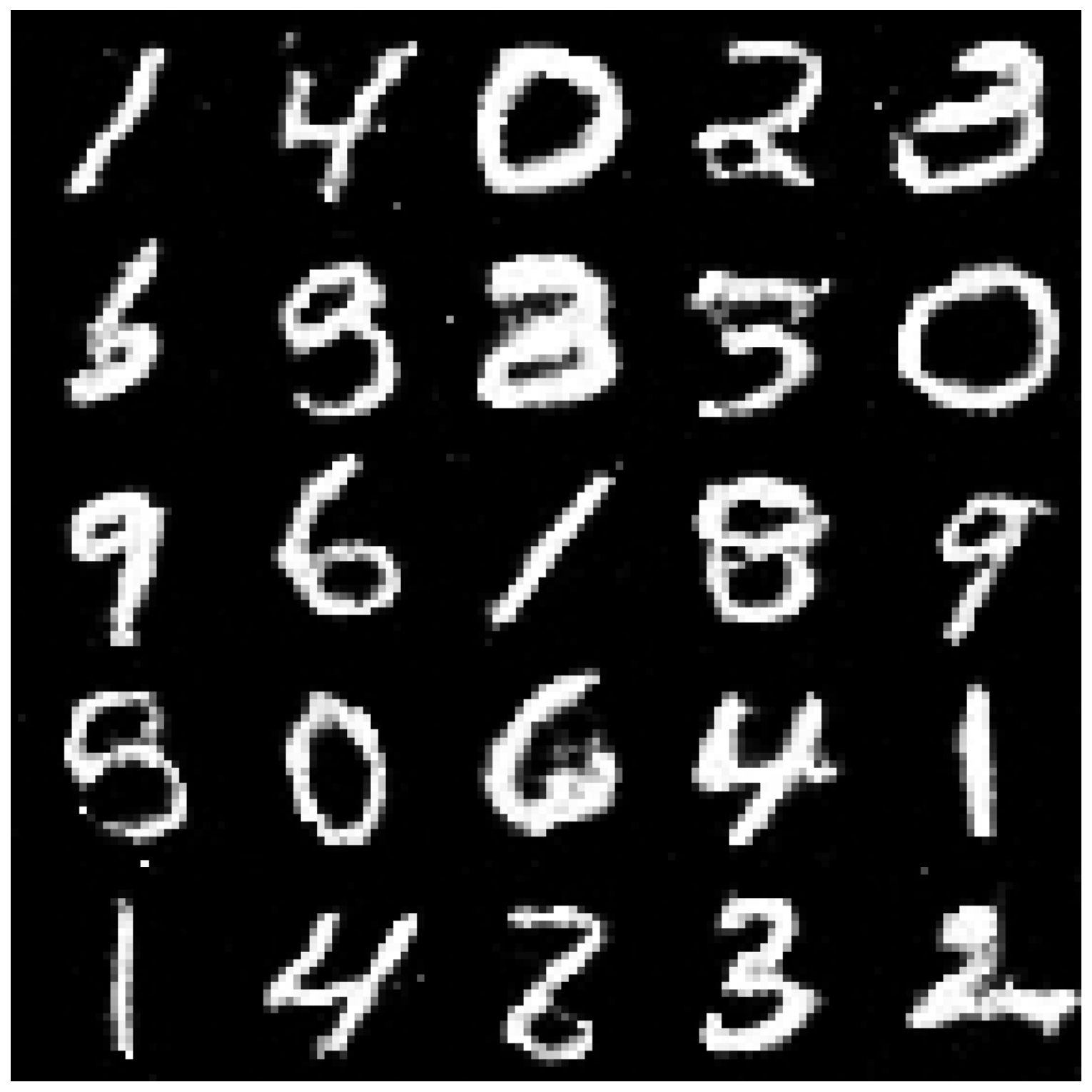}
    \includegraphics[width=0.12\textwidth]{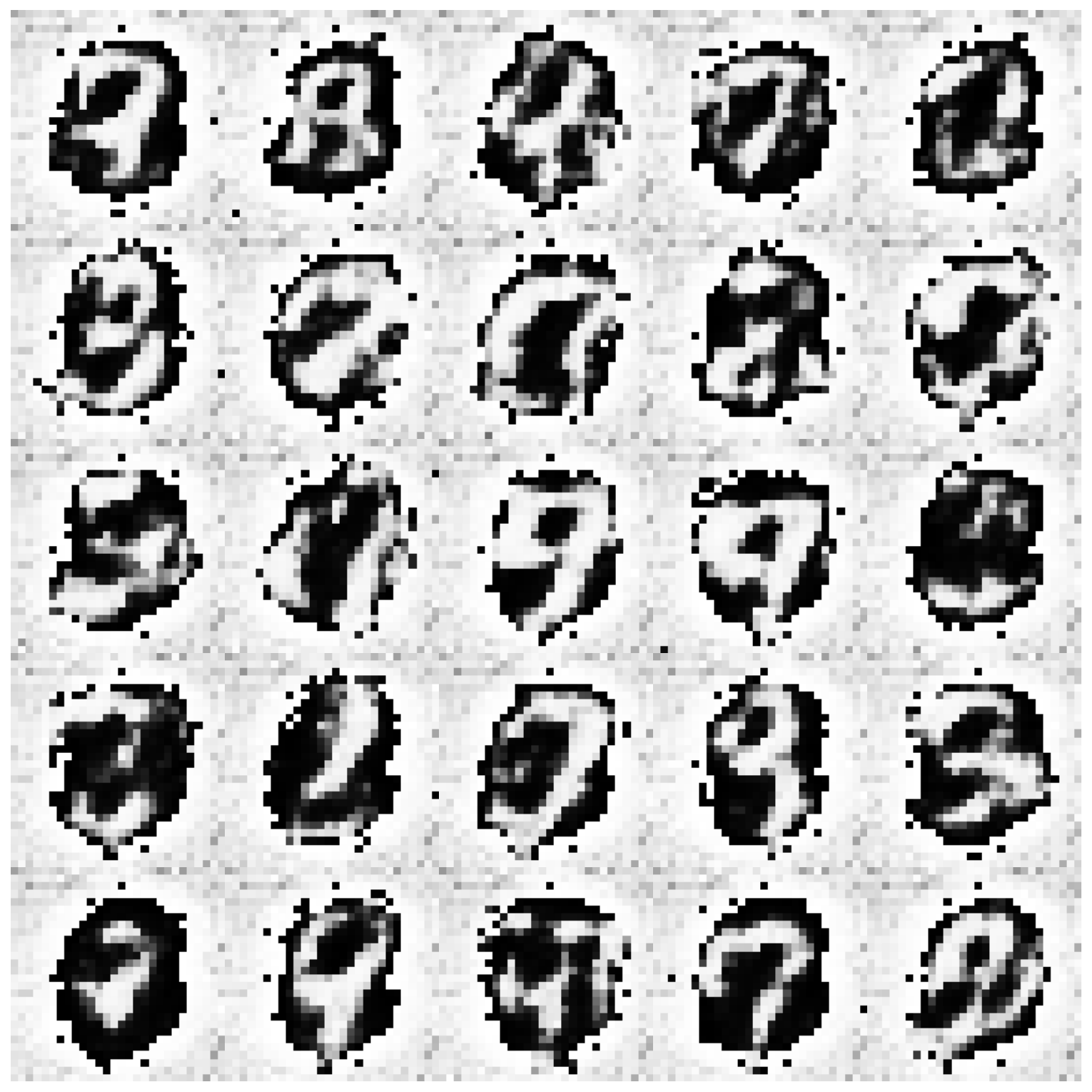}
    \includegraphics[width=0.12\textwidth]{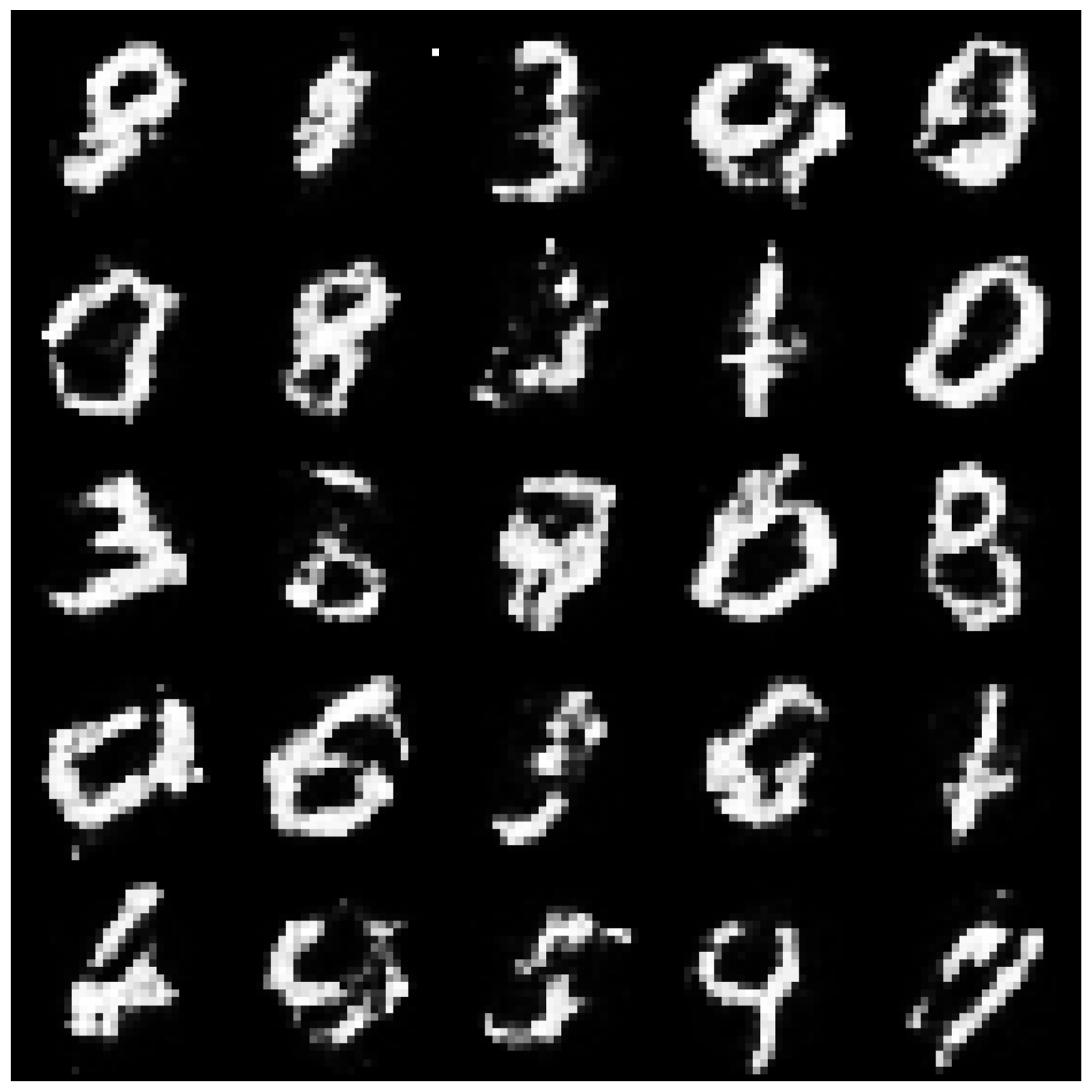} 
    \includegraphics[width=0.12\textwidth]{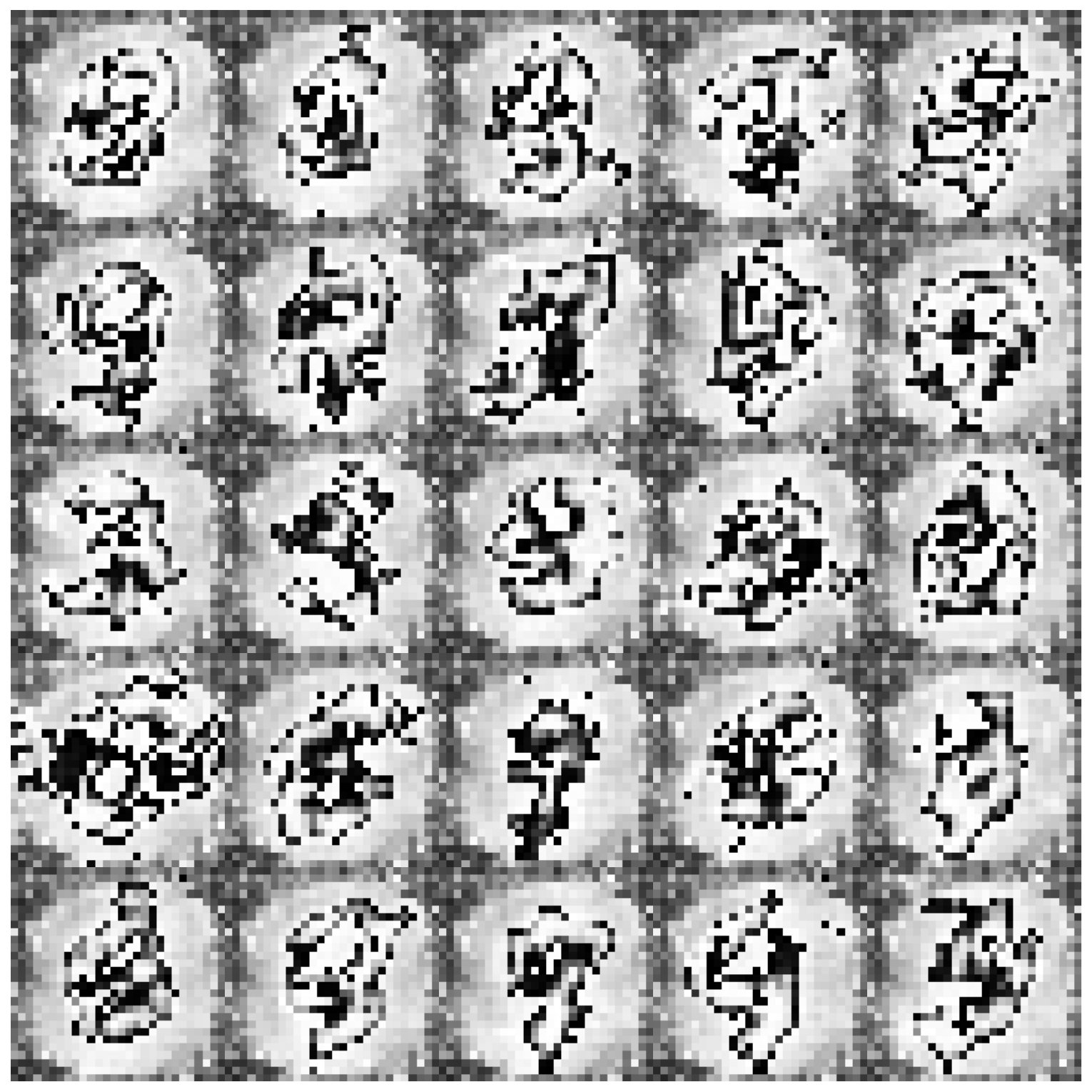} 
    \includegraphics[width=0.12\textwidth]{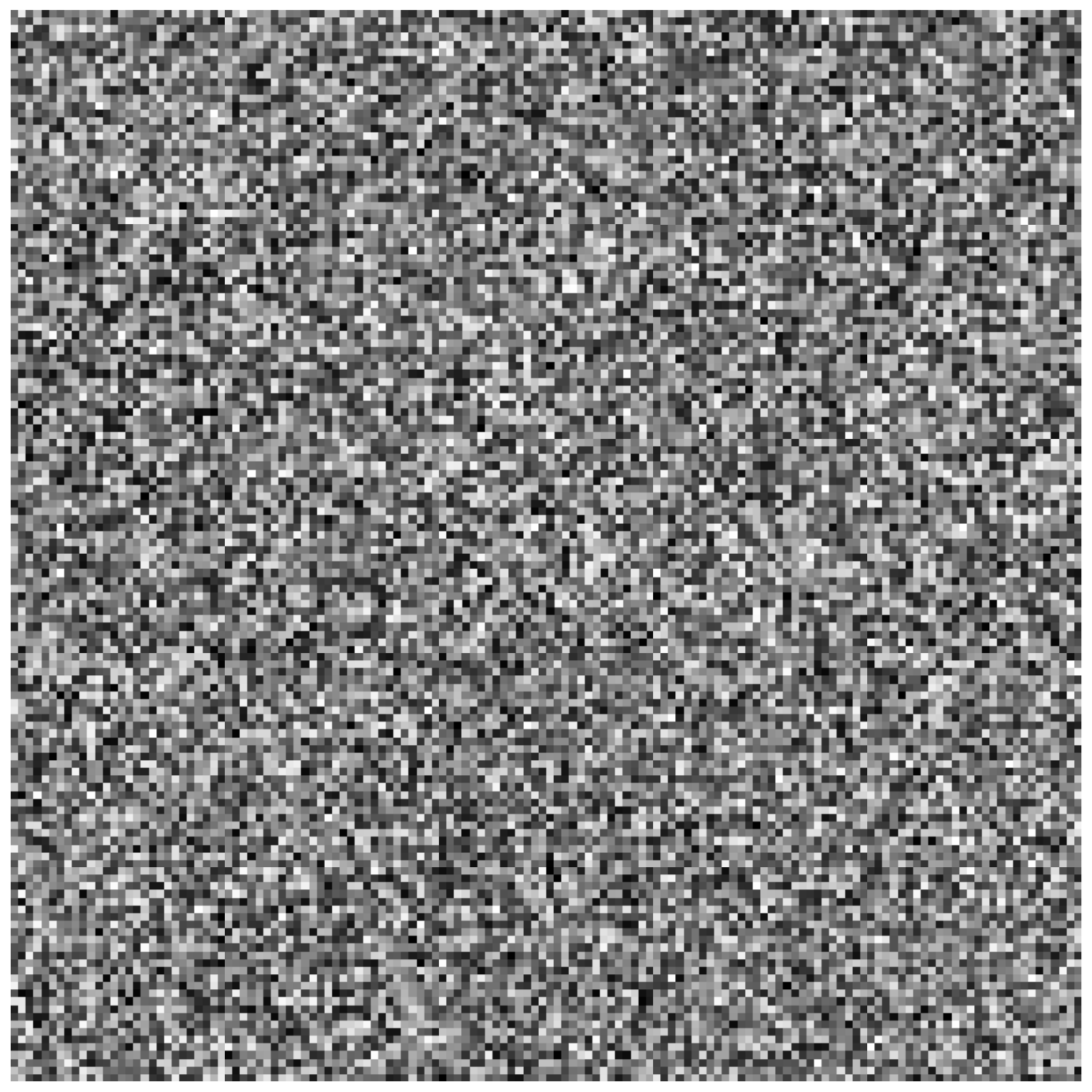} 
    \includegraphics[width=0.12\textwidth]{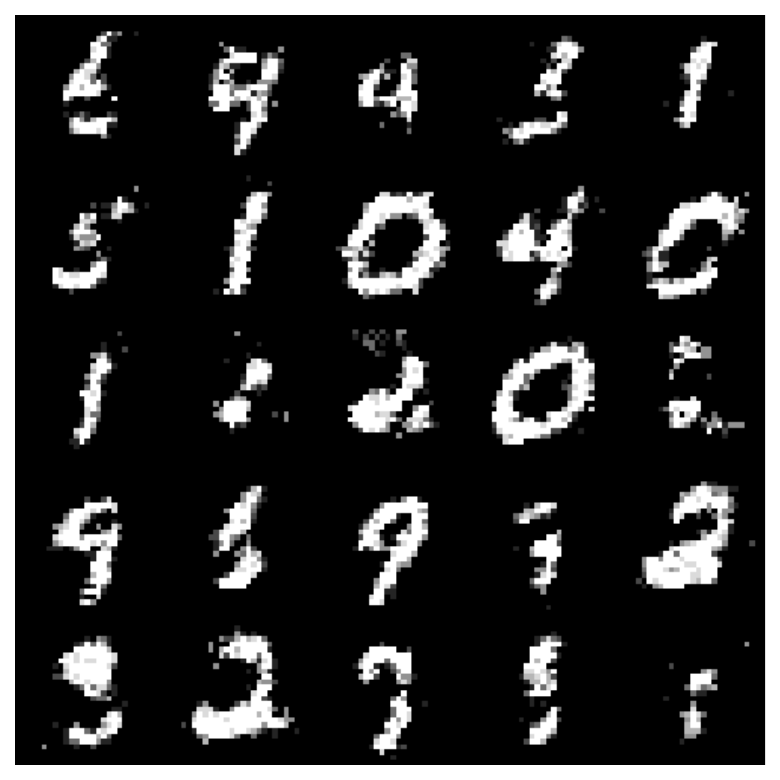} 
    \includegraphics[width=0.12\textwidth]{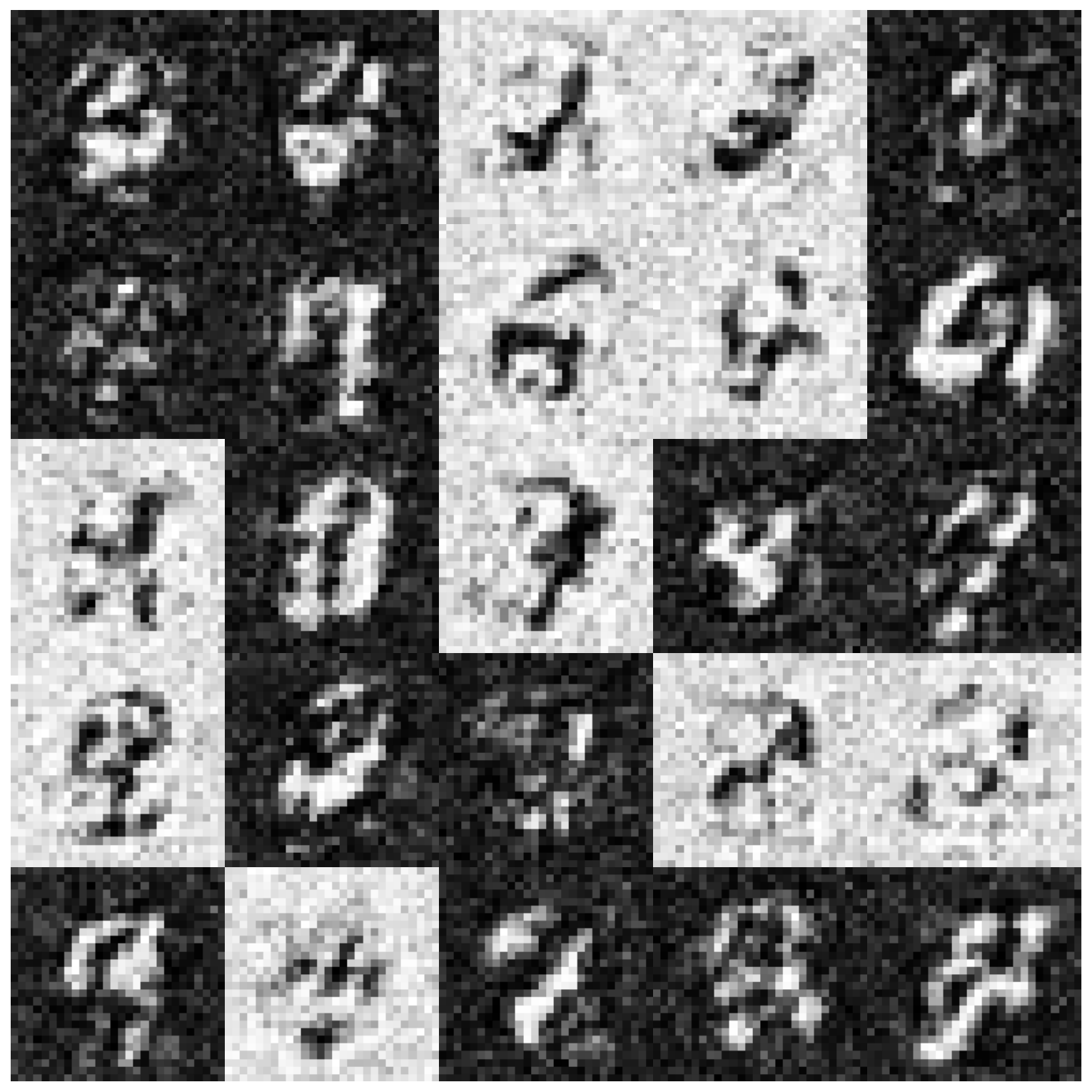} \\ 
    
    \includegraphics[width=0.12\textwidth]{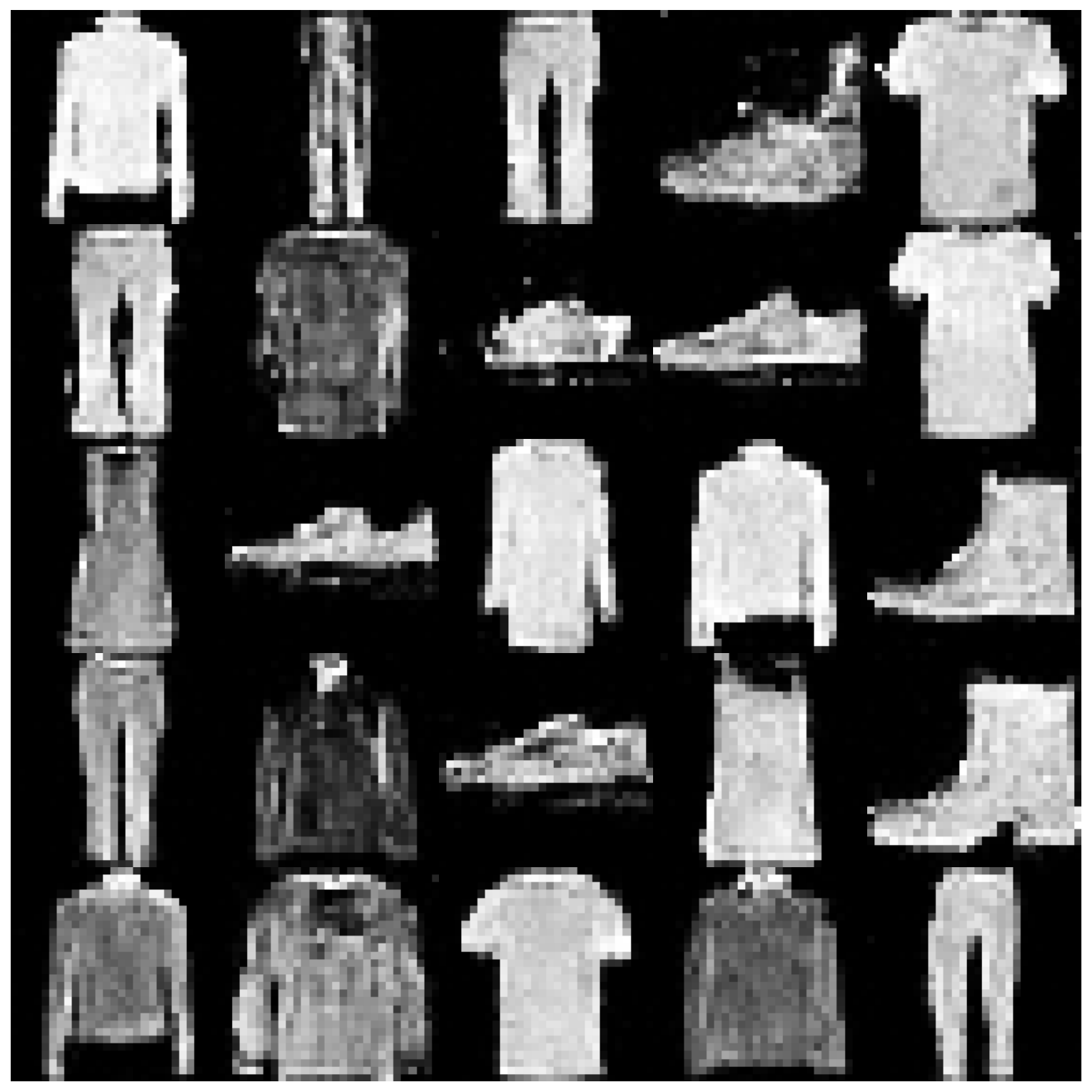} 
    \includegraphics[width=0.12\textwidth]{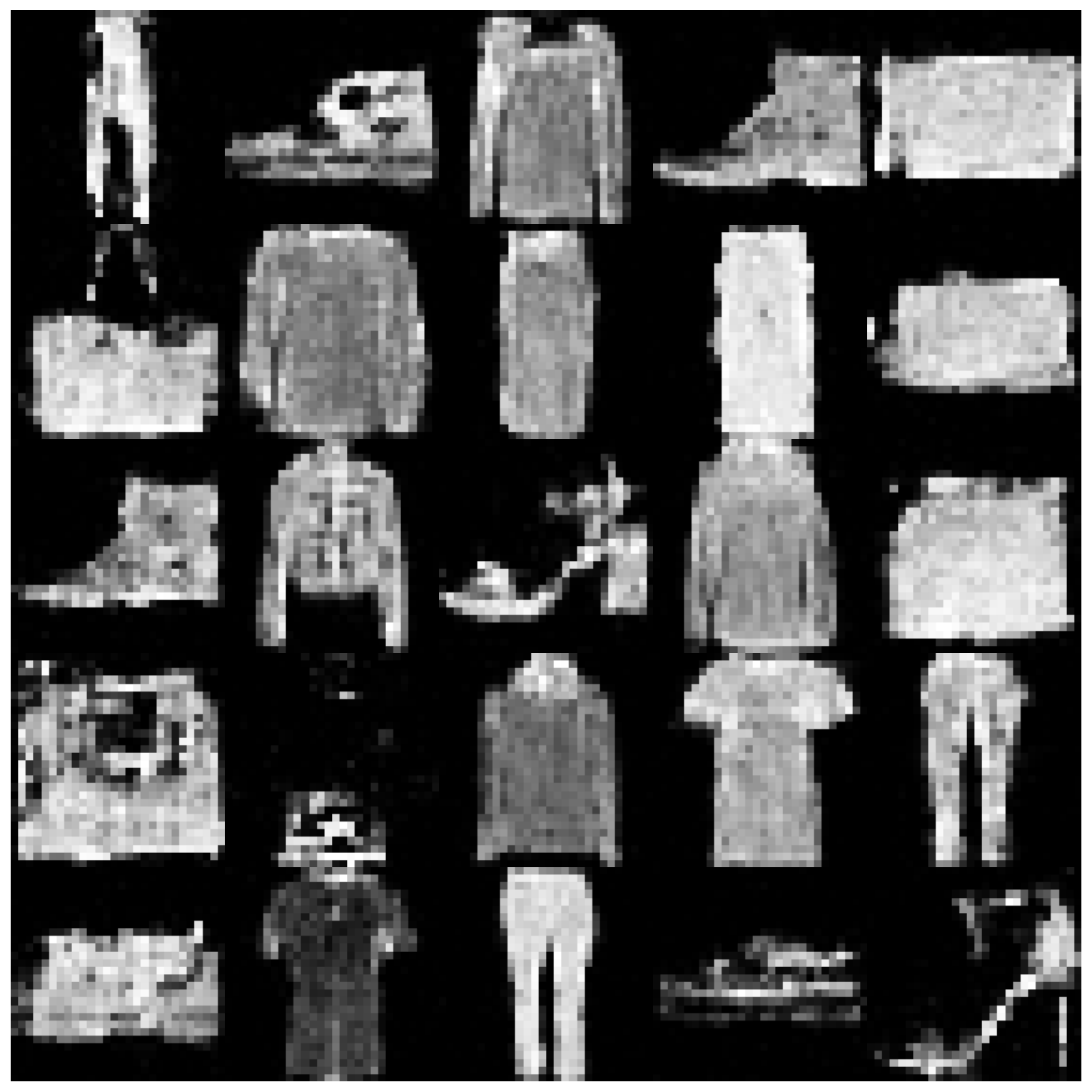}
    \includegraphics[width=0.12\textwidth]{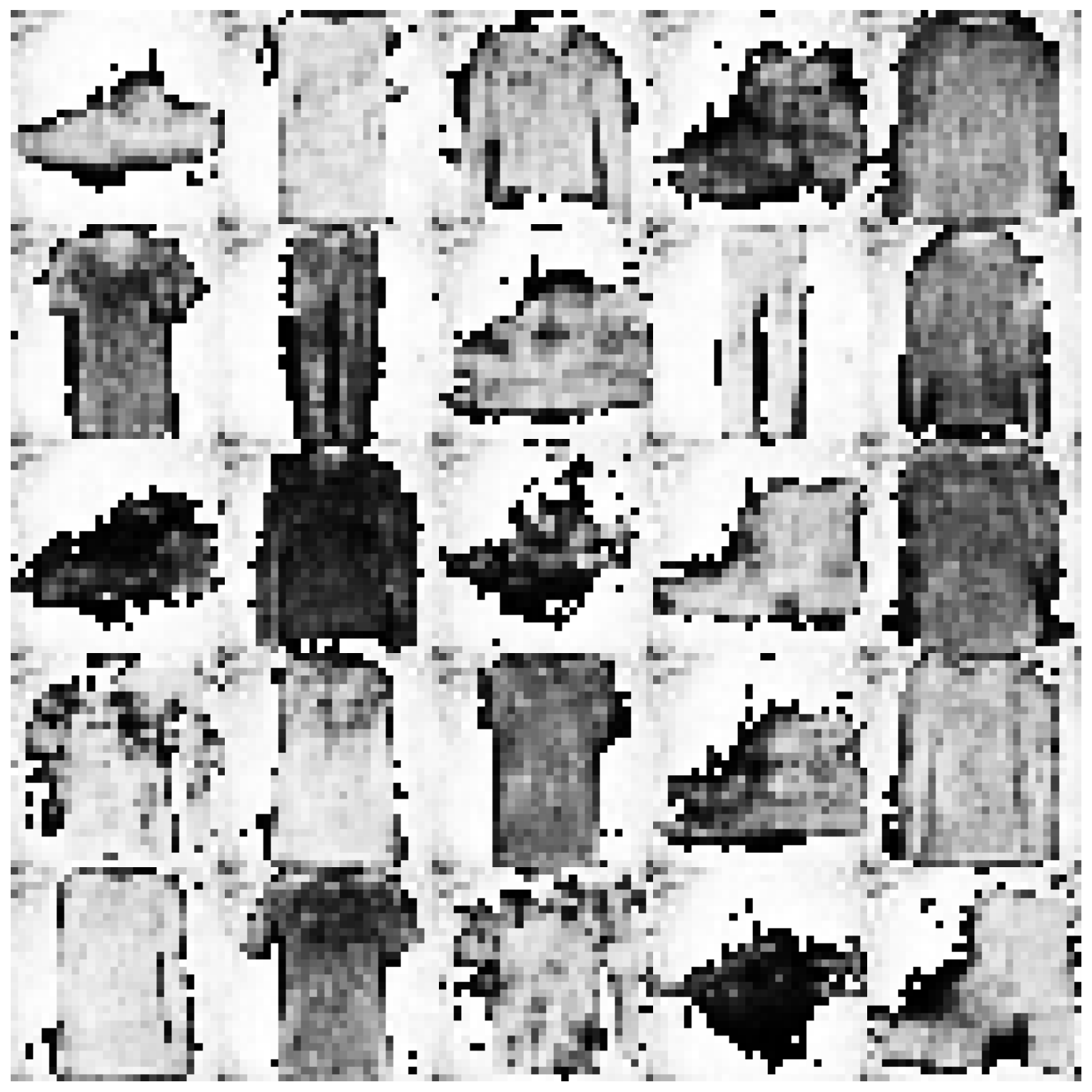}
    \includegraphics[width=0.12\textwidth]{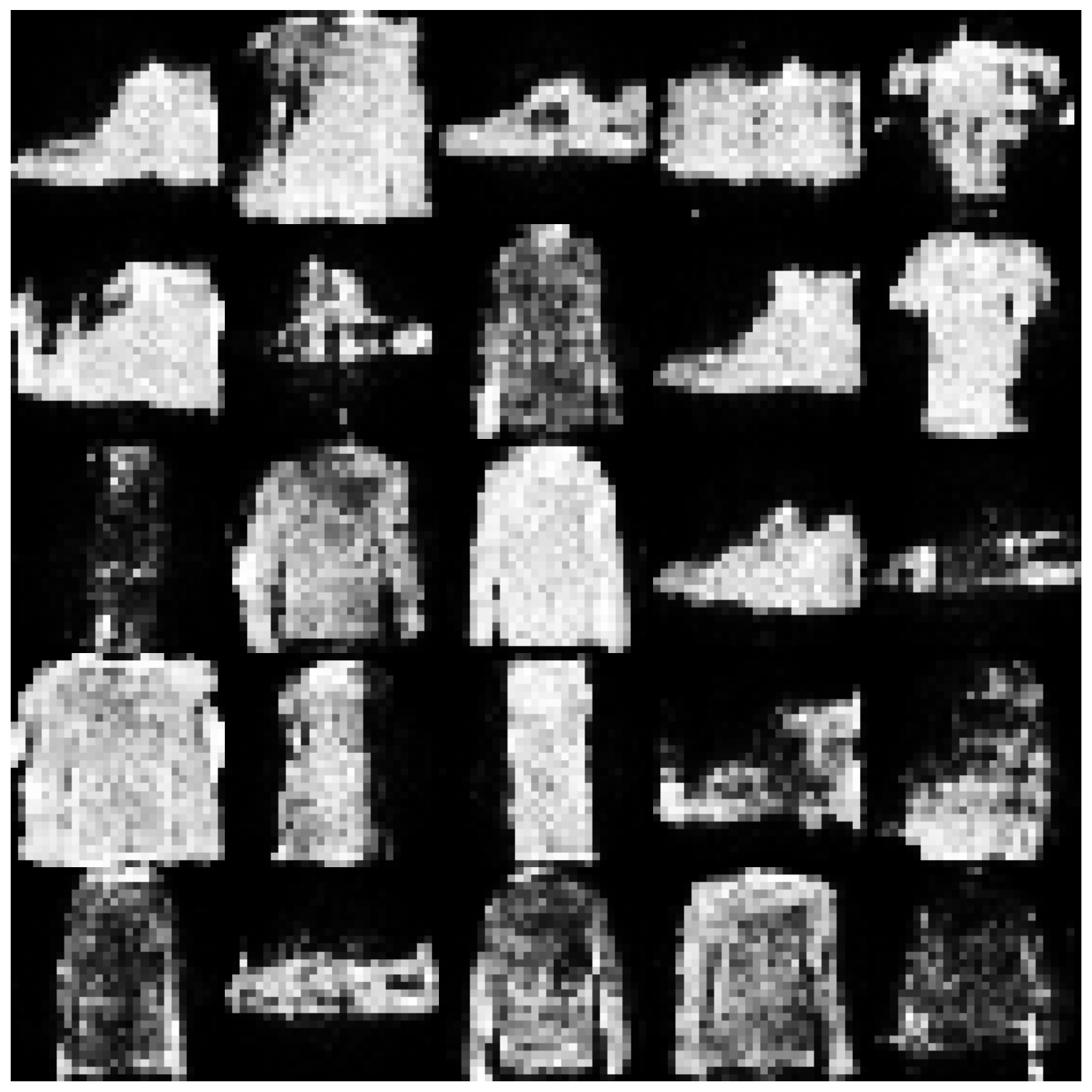} 
    \includegraphics[width=0.12\textwidth]{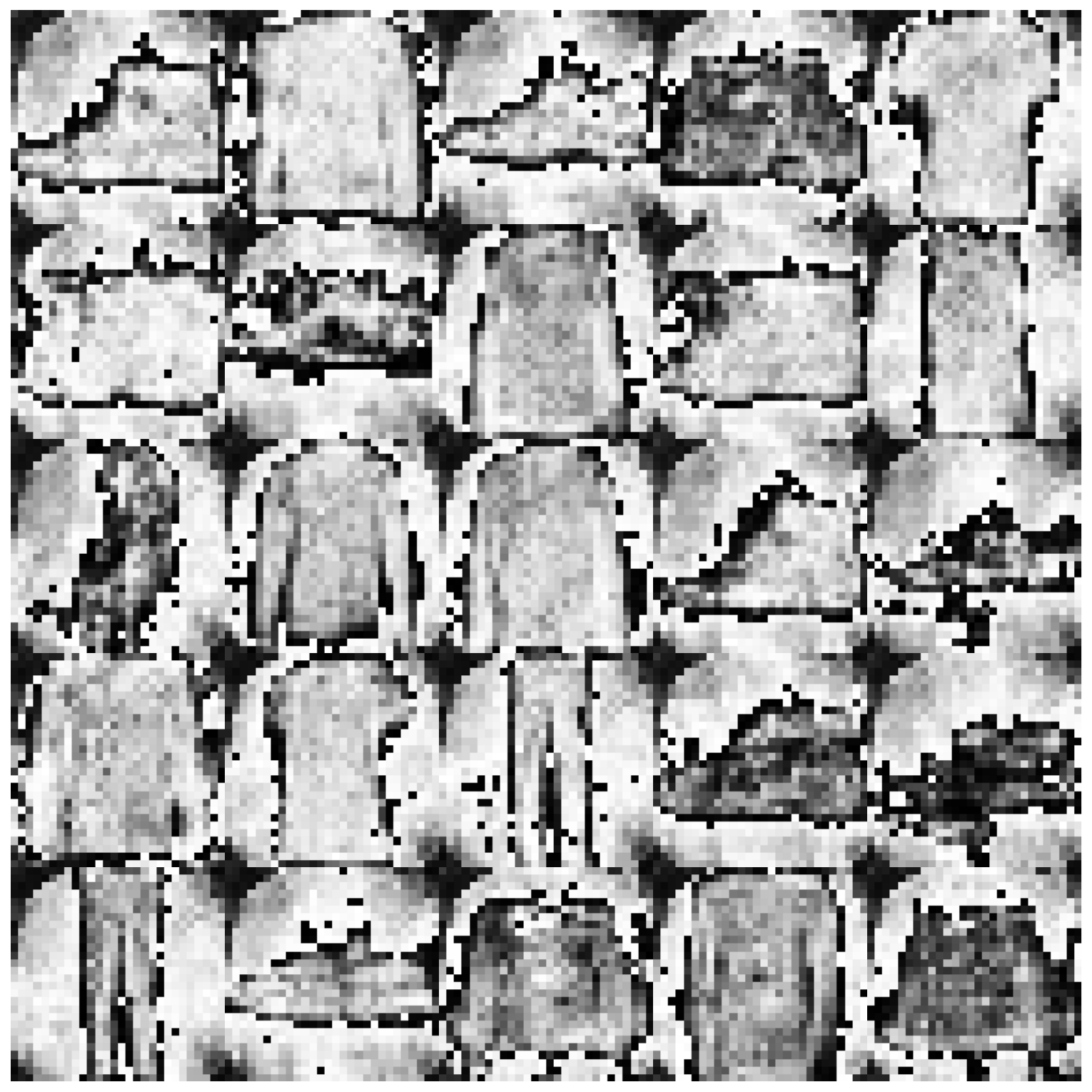} 
    \includegraphics[width=0.12\textwidth]{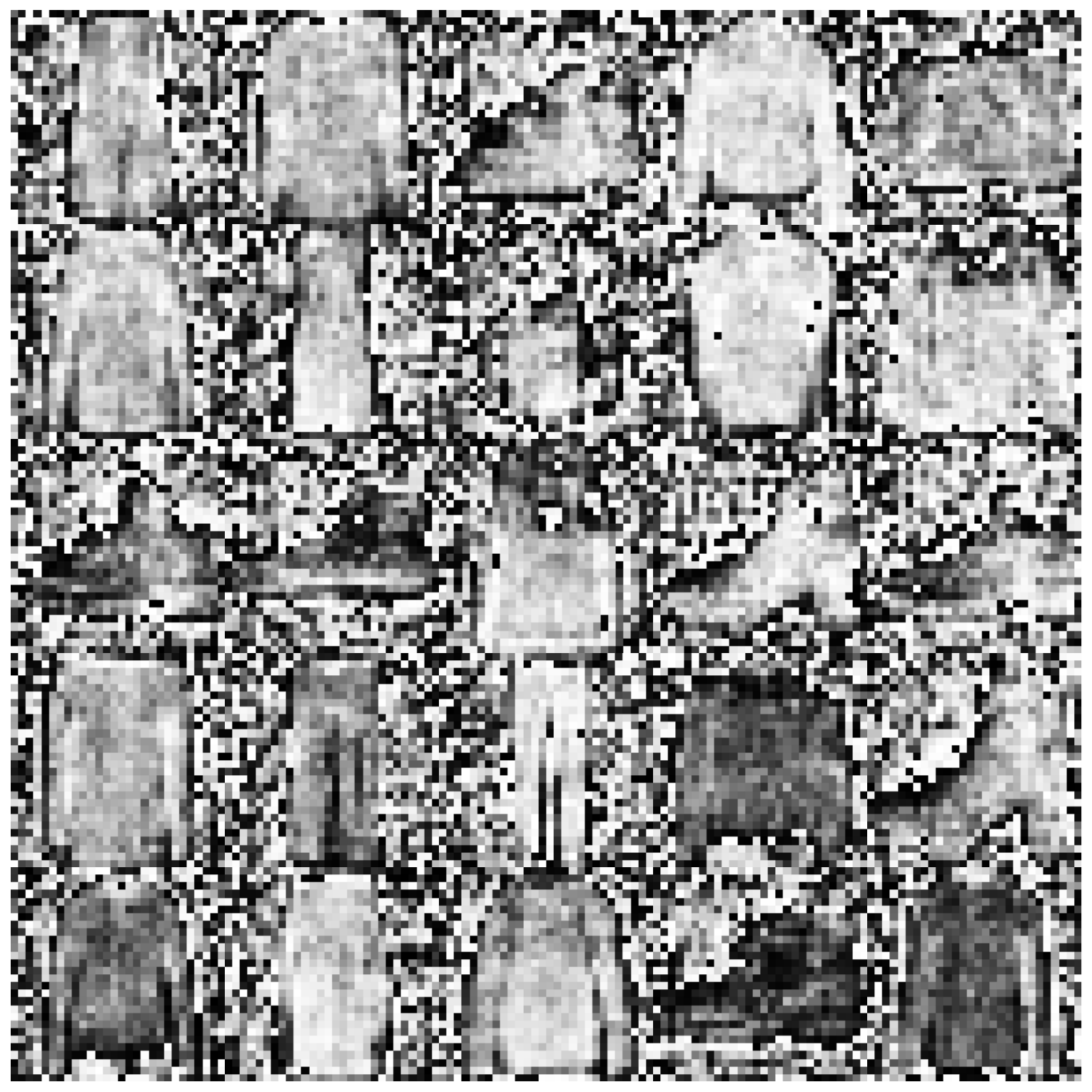} 
    \includegraphics[width=0.12\textwidth]{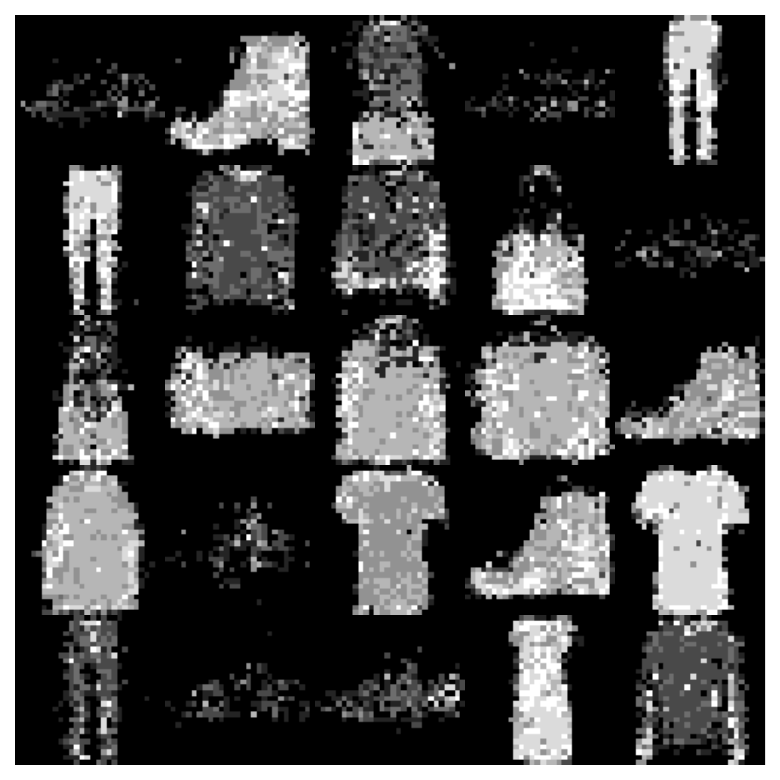} 
    \includegraphics[width=0.12\textwidth]{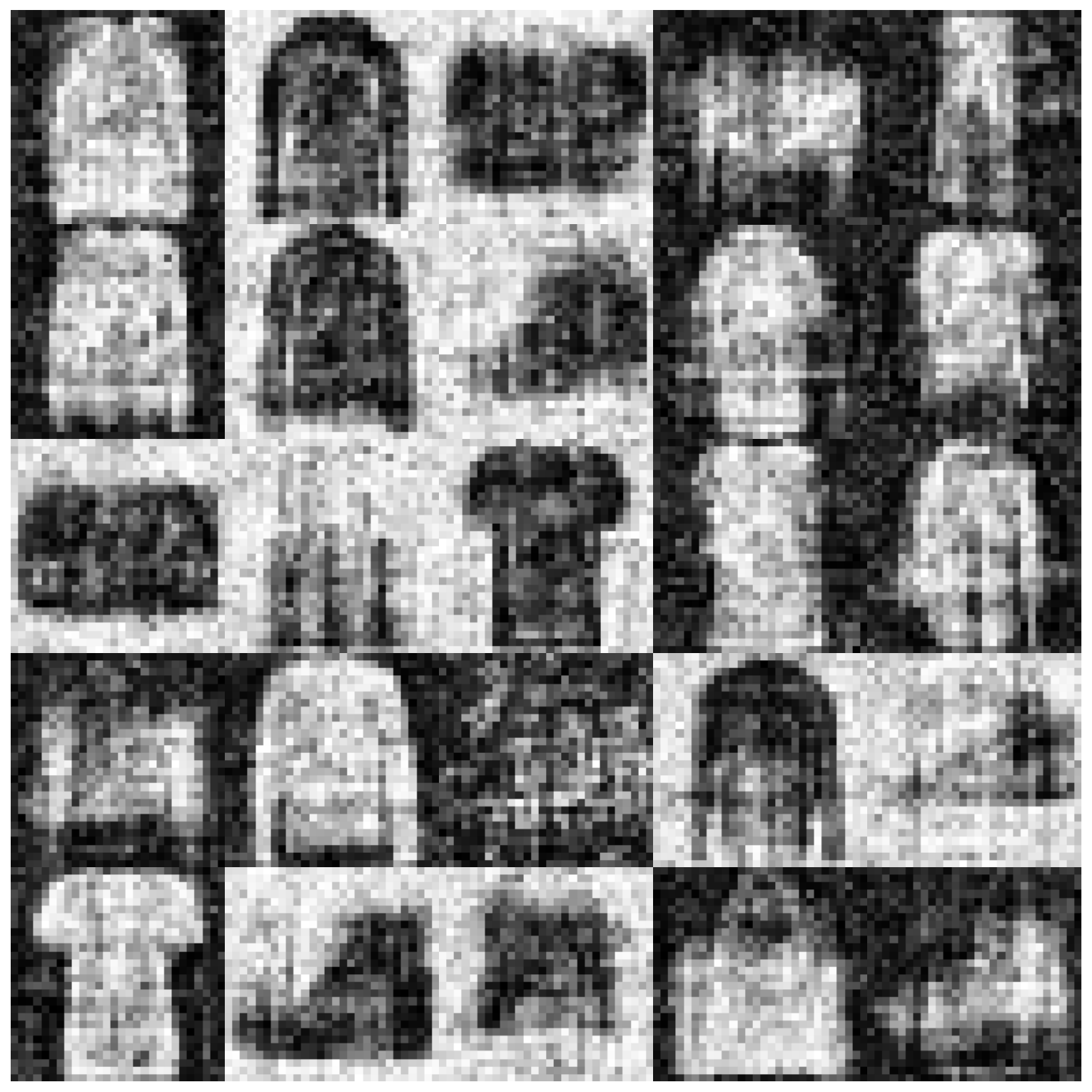}
    
    \caption{Generated MNIST/FMNIST samples. KS-Nonideal: Kuramoto-SHIL with $\weightbit=4$ weights and noise $\noiseamp=0.025$.}
    \label{fig:visual_comparison}
\end{figure*}

\begin{figure}
    \centering
    \includegraphics[width=0.9\linewidth]{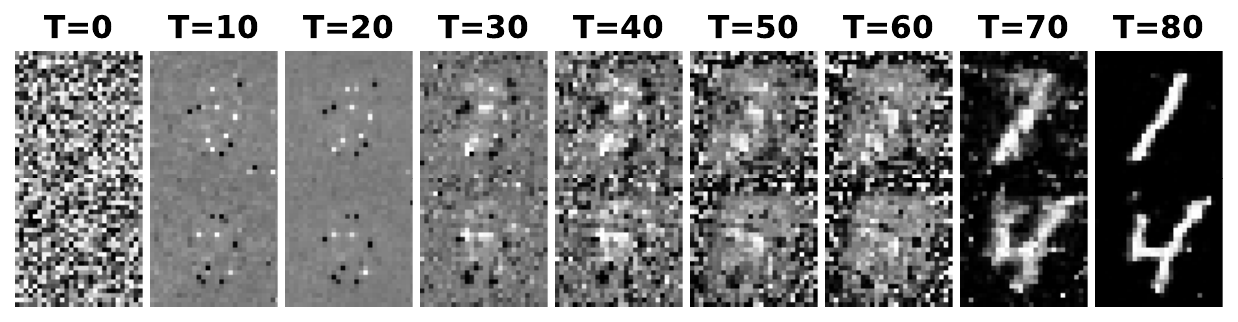}
    \caption{Images generation with KuraSHIL.}
    \label{fig:state-evol}
\end{figure}

We first conduct a design-space exploration (DSE) of the \modelabbr{} architecture, evaluating coupling topologies and analog dynamics under simulated hardware constraints (weight quantization and noise). We assume the time-varying coupling is implemented in a state-stationary manner (Figure~\ref{fig:state_stat}).
We solve the dynamical system using Euler's method with 80 steps of size $0.05$ and 64-bit floating-point precision. We sample initial states from $\mathcal{N}(0, 0.25)$. 
For the discriminator, we employ a DCGAN architecture~\cite{radford2015dcgan}. 
The system is trained for 30,000 generator steps with a batch size of 64 using the Adam optimizer with $(\beta_1, \beta_2)=(0.5, 0.9)$. 
The learning rate decays linearly from $10^{-4}$ to zero during the second half of training. 
We fine-tune each configuration for 15,000 additional steps using the same schedule. 
We report the average Fréchet Inception Distance (FID) over four random seeds on the MNIST and Fashion-MNIST (FMNIST) datasets~\cite{lecun2002mnist,xiao2017fmnist}.
Table~\ref{tab:resource_config} reports the resource utilization.

\subsection{Comparison with Baselines} 
The development of hardware-implementable generative models, where computation emerges naturally from the physical interactions of coupled elements, is still in its nascency, and training these systems remains challenging. Therefore, the literature contains few comparable models. We benchmark \modelabbr{} against recent state-of-the-art frameworks in the field, specifically the Denoising Thermodynamic Model (DTM)~\cite{jelinvcivc2025dtm} and the Neural Langevin Machine (NLM)~\cite{yu2025nlm}. 
The DTM consists of a chain of Ising solvers to execute diffusion. Each solver uses a probabilistic-bit (p-bit) grid that updates based on local interactions. 
The NLM consists of a continuous dynamical system with a specialized learning rule designed to minimize the Kullback-Leibler divergence between the state distribution and the target data. Because neither work reports FID scores on grayscale images, we reproduce their setups using their official source code. 

We configure the DTM with an $80 \times 80$ grid, 24 couplings per node, and 4 diffusion steps, yielding an iso-parameter comparison with \modelabbr{}. Grayscale pixels are binary-encoded using three p-bits each. The DTM is trained for 60 epochs across four seeds, ensuring loss convergence. 
We follow the original NLM training procedure and verify loss convergence. For DTM, we validate that we observe better FID scores than what is reported in their logs.

\prh{Parameter Counts.} Table~\ref{tab:resource_config} summarizes the resource utilization for each model. Because the DTM executes diffusion, it requires multiple copies of its Ising solver, which quadruples the number of PEs and couplings. This cost can be amortized by pipelining. However, while the DTM logically employs symmetric coupling, its physical implementation relies on unidirectional resistor networks. This architectural choice doubles the total number of hardware couplings required to realize bidirectional symmetric weights. The NLM requires dense, all-to-all connectivity and does not utilize hidden states. As a result, even though the NLM operates with fewer state variables overall, it demands twice the parameter count of \modelabbr{}.

\prh{Fidelity} The fidelity comparison is presented in Figure~\ref{fig:visual_comparison}. DTM achieves FIDs of 107.8/112.8 and NLM achieves 230.5/200.8 on MNIST/FMNIST, respectively. Our model substantially outperforms both, achieving FIDs of 27.6/80.8 on MNIST/FMNIST at full precision with no noise, and 29.1/75.1 with 4-bit quantization and $\noiseamp=0.025$. Qualitative comparisons of the generated samples are provided in Figure~\ref{fig:visual_comparison}. Notably, even when subjected to hardware constraints (4-bit weights and transient noise), our model consistently generates higher-quality images using an equivalent resource footprint.

\subsection{Design-Space Exploration}
To evaluate the effect of different design decisions, we sweep multiple design points, detailing the results in Table~\ref{tab:dse}. We initially assume full-precision weights and noiseless transients to identify promising architectures, subsequently introducing quantization and noise to evaluate hardware robustness.

\prh{Analog Models.} We perform initial experiments with \modelabbr{} models in Table~\ref{tab:models}, configuring all with asymmetric and time-varying ($\temporalweight=4$) coupling. KuraSHIL achieves substantially lower FID scores on both datasets compared to the other \modelabbr{} models (27.6/80.8 vs. 307.3/160.3, 90.7/95.4, 288.77/144.82, and 335.76/237.11 on MNIST/FMNIST, respectively). This advantage is corroborated by qualitative visual improvements, as shown in Figure~\ref{fig:visual_comparison}. Consequently, we select the KuraSHIL model as our primary analog compute primitive for all subsequent evaluations.

\prh{Temporal and Asymmetric Weights.} Increasing the number of temporal weights from 1 to 4 consistently reduces the FID scores across both datasets. This highlights that time-varying weights enhance the model's expressivity, even when limited to piecewise-constant functions. 
Similarly, employing asymmetric coupling introduces additional degrees of freedom in the parameter space, enabling the model to learn more complex distributions and generate lower-FID images compared to symmetric constraints.

\prh{Hidden States} We run an experiment with a grid of size $28\times28$ with $\temporalweight=4$, having only visible states. The FID score increases for both datasets (27.6/80.8 to 43.0/84.3). This demonstrates that adding hidden states increases the expressivity of an \modelabbr{} model under sparse connection constraints.

\prh{Quantization and Noise.} We evaluate the model's resilience to hardware non-idealities by varying weight precision and noise levels on the 4-temporal-weight configuration. To train the quantized models, we initialize them with hard-quantized weights derived from the full-precision baseline prior to fine-tuning. We subsequently inject varying magnitudes of noise and further fine-tune these quantized models. Both fine-tuning stages involve 15,000 generator iterations. 

The 4-bit quantized model achieves performance comparable to the 64-bit unquantized baseline (27.2/81.9 vs. 27.6/80.8 FID on MNIST/FMNIST). However, further reducing precision noticeably degrades generation quality. Introducing minor transient noise ($\noiseamp=0.025$) to the 4-bit model yields negligible degradation and even shows a slight improvement on FMNIST. As expected, aggressively increasing the noise level from 0.025 to 0.1 causes a steady decline in quality, raising the FID from 29.1 to 54.5 on MNIST and from 75.1 to 90.4 on FMNIST. Overall, the \modelabbr{} demonstrates resilience to realistic hardware constraints.

The generation process is visualized in Figure~\ref{fig:state-evol}. Interestingly, the oscillators do not learn a direct interpolation path from the initial Gaussian noise to the final images. Instead, they learn a detoured trajectory that initially converges toward average pixel values before spreading out into distinct features. This behavior demonstrates the effectiveness of our GAN-based training formulation in discovering non-trivial dynamical paths that successfully map the Gaussian prior to the target data distribution.



\section{Related Work}\label{sec:related}

\prh{Oscillatory Neural Networks}
Recent works leverage oscillatory dynamics to overcome limitations in standard neural networks. GraphCON and SLGNN apply second-order oscillatory ODEs to mitigate oversmoothing and enrich representations in graph neural networks~\cite{rusch2022graphcon, zhang2025slgnn}. AKORN and Kuramoto Orientation Diffusion utilize Kuramoto phase synchronization to improve feature binding and robustness, showing success on image processing and generative modeling~\cite{miyato2024akorn, song2025kuramoto-diffusion}. While these prior works inject oscillatory dynamics into standard neural networks, they still rely on unconstrained digital computing and are not realizable on analog hardware. Our goal is fundamentally different: we strictly constrain computation to \textit{only} use physical oscillatory or specific analog interaction systems with bounded hardware limits to enable physically realizable analog architecture for low-power, low-latency compute.

\prh{Generative Models with Thermodynamic Computing.}
Harnessing the physical evolution of natural systems for computation is broadly categorized as thermodynamic computing~\cite{conte2019thermodynamic-computing}. A prominent direction in this domain relies on Ising models implemented via probabilistic bits (p-bits). Recent implementations include sparse Ising machines synthesized on FPGAs for image tasks~\cite{niazi2024sparse-ising-rbm} and Denoising Thermodynamic Models (DTMs) that chain multiple Ising solvers to compute discrete diffusion steps~\cite{jelinvcivc2025dtm}. While these architectures achieve low-power generation for binarized images, scaling them to multi-bit data (e.g., grayscale images) introduces noise due to the discrete integer encoding, as evaluated in~\cite{jelinvcivc2025dtm} and Section~\ref{sec:eval}. The optimal embedding of continuous variables onto discrete p-bit hardware remains an unresolved challenge.

Alternative approaches employ continuous physical dynamics~\cite{yu2025nlm, whitelam2026generative-thermodynamic-computing}. This aligns closely with our formulation, enabling the direct mapping of continuous data values to physical states. However, existing training methodologies for these continuous systems yield substantially higher FID scores (Section~\ref{sec:eval}). We overcome the optimization bottleneck by identifying the GAN formulation as an effective training paradigm for physical dynamics.

\prh{Generative Models with Non-Digital Neural Networks}
Recent efforts map generative models onto non-digital substrates, such as optical systems~\cite{chu2026adm, qiu2025optical-gan, oguz2024optical-diffusion, chen2025optical-generative-model} and emerging memory arrays~\cite{cheng2025MRAM-diffusion, yang2024RRAM-diffusion}, demonstrating improvements in energy efficiency. These architectures operate by emulating pre-defined digital algorithms, such as digital score-based diffusion processes parameterized by standard neural networks, and utilize the non-digital fabric as a drop-in accelerator for specific operations, such as matrix-vector multiplication.

Our Analog Interaction System follows a different philosophy. We formulate the generative process natively within the hardware's physics, rather than forcing analog circuits to emulate a digital neural network. By employing a GAN framework to optimize the terminal state of this physical system, we eliminate the requirement to track a predefined digital trajectory, which is often too complicated to fit for physical systems.

\prh{Analog Accelerators.} 
Extensive research has been dedicated to developing analog accelerators, particularly oscillator-based Ising machines, engineered primarily to solve complex combinatorial optimization problems such as Boolean satisfiability (SAT) and graph coloring~\cite{moy2022-1968node-con, ahmed2021con-neighbor, mallick2021con-global, graber2024oim-local-dac}. A central focus within this community involves navigating the trade-offs in hardware topology and component design. Our work is complementary to these hardware endeavors. \modelabbr{} opens a new application domain by leveraging the continuous-time dynamics for generative AI, while physical innovations in circuit design provide the substrate to scale and implement the \modelabbr{}.

\section{Conclusion}\label{sec:conclusion}

Analog hardware offers a compelling path toward energy-efficient generative modeling, but realizing this potential requires bridging a fundamental mismatch between the fixed, physics-determined dynamics of analog substrates and the flexibility demanded by modern generative models. 
We have introduced Analog Interaction Systems as a unified framework for reasoning about this class of hardware, and identified two complementary mechanisms — hidden state augmentation and time-varying weights — that reduce the expressivity gap while remaining compatible with physical implementation constraints. 
To train these systems, we applied a Wasserstein GAN formulation that frees the underlying physics to discover trajectories from noise to data. We further proposed a sparse, low-bit-width hardware architecture and provided scaling analyses showing that such a design can yield order-of-magnitude energy improvements over digital alternatives.
Taken together, these results suggest that the limitations imposed by analog hardware can be overcome to yield expressive, energy-efficient models.


\bibliographystyle{ACM-Reference-Format}
\bibliography{references}

\end{document}